\documentclass[rmp,twocolumn,letterpaper]{revtex4}
\makeatletter
\def\raggedcolumn@skip{\vskip\z@\@plus.0001fil\relax}\makeatother
\usepackage[bookmarks=false]{hyperref}
\usepackage{orcidlink,graphicx,newtxtext,newtxmath}
\def\eprint#1{E-print
\href{https://arxiv.org/abs/#1}{\nolinkurl{arXiv:#1}}}
\def\doi#1{\href{https://doi.org/#1}{\nolinkurl{doi:#1}}}
\hypersetup
{pdftitle={Jacobi's solution for geodesics on a triaxial ellipsoid},
 pdfauthor={C. F. F. Karney}}
\date{\today}

\usepackage{array}
\def\abs#1{\left|#1\right|}
\def\round#1{\lfloor#1\rceil}
\DeclareMathOperator{\sign}{sign}
\DeclareMathOperator{\am}{am}
\DeclareMathOperator{\cn}{cn}

\DeclareMathOperator{\dn}{dn}
\DeclareMathOperator{\tr}{tr}
\newcommand{\atanx}[2]{\tan^{-1}\genfrac{}{}{1.4pt}{0}{#1}{#2}}
\newcommand{\half}{{\textstyle\frac12}}
\renewcommand{\d}{\mathrm d}
\def\figuredir{figures}
\begin{document}
\title{Jacobi's solution for geodesics on a triaxial ellipsoid}

\author{Charles F. F. Karney\,\orcidlink
  {0000-0002-5006-5836}}
\email[Email addresses: ]{charles.karney@sri.com}
\thanks{\href
  {mailto:karney@alum.mit.edu}{karney@alum.mit.edu}.}
\affiliation{\href{https://www.sri.com}{SRI International},
201 Washington Rd, Princeton, NJ 08540-6449, USA}

\begin{abstract}
On Boxing Day, 1838, Jacobi found a solution to the problem of geodesics
on a triaxial ellipsoid, with the course of the geodesic and the
distance along it given in terms of one-dimensional integrals.  Here, a
numerical implementation of this solution is described.  This entails
accurately evaluating the integrals and solving the resulting coupled
system of equations.  The inverse problem, finding the shortest path
between two points on the ellipsoid, can then be solved using a similar
method as for biaxial ellipsoids.
\end{abstract}
\maketitle

\section{Introduction}

The shortest path between two points on a surface is a geodesic, and it
plays a crucial role in geodesy, where the earth is typically modeled as
an ellipsoid of revolution, a biaxial ellipsoid.  The main geodesic
problems are (1) given a starting point and a direction, find the point
a certain distance away, the ``direct'' geodesic problem, and (2) to
find the length and direction of the geodesic connecting two points, the
``inverse'' geodesic problem.  The path of a geodesic is also given by a
point mass constrained to the surface of the ellipsoid and moving
without friction and in the absence of external forces---this path is
``as straight as possible.''  This allows geodesics to be extended
indefinitely.

Because of their central role in geodesy, there is naturally an interest
in exploring the behavior of geodesics when the earth is modeled as a
triaxial ellipsoid \citep{bursa93}.  In addition, there are other bodies
in our solar system which are not rotationally symmetric but which can
be approximated by a triaxial ellipsoid.  Consequently, reliable
solutions to the geodesic problems for a triaxial ellipsoid would be
useful.

The solution of the direct geodesic problem in the biaxial case is
relatively straightforward.  The angular momentum about the axis of
symmetry (the Clairaut constant) is conserved, allowing the path to be
found in terms of elliptic integrals.

The triaxial ellipsoid, on the other hand, possesses no obvious
symmetry.  It therefore came as a surprise when \citet{jacobi39} found
that the geodesic problem could be reduced to quadrature in this case
too, with the solution given in terms of one-dimensional integrals.  (We
know the date, given in the abstract, for this discovery, because of a
letter he wrote two days later to F.~W. Bessel, his neighbor in
K\"onigsberg.)
\citet[\S28]{jacobi43} expanded on his method in his {\it Lectures on
Dynamics}, and the result was generalized
by \citet[\S\S20--21]{liouville46b} to apply to so-called Liouville
surfaces.  The qualitative properties of the solution can be found in
several
textbooks \citep{darboux94,hilbert52,klingenberg82,arnold89,berger10}.
However, unlike the case of the biaxial ellipsoid,
where \citet{bessel25-en} provided a prescription for computing
geodesics, little effort was given to {\it implementing} Jacobi's
solution.

The goal of this paper is to address this deficiency, specifically to
enable the direct and inverse problems to be solved with high accuracy
and reasonable efficiency.  This entails approximating the integrands as
Fourier series, which allows the indefinite integrals to be easily
evaluated, and finding an efficient way to solve the resulting coupled
system of equations.  With the solution to the direct problem in hand,
we turn to solving the inverse problem, following the same basic recipe
used in the biaxial case \citep{karney13,karney-geod2}.

\citet{panou13,panou19} explore an alternative approach to
solving the direct geodesic problem, namely by numerically integrating
the corresponding ordinary differential equations for the geodesics, as
discussed in Appendix \ref{app-ode}.  This can provide an accurate
solution, although the properties of the true solution are only
approximately maintained.  They do not provide a complete solution to
the inverse problem.

\section{Ellipsoidal coordinates}

Jacobi's insight was to express the equations of the geodesic in terms
of {\it ellipsoidal} coordinates; this allows the equations to be
reduced to one-dimensional integrals through the separation of
variables.

Consider the ellipsoid defined by
\begin{equation}
  S(\mathbf R) = \frac{X^2}{a^2} + \frac{Y^2}{b^2} + \frac{Z^2}{c^2} - 1 = 0,
\end{equation}
where $\mathbf R = [X, Y, Z]^T$ is a three-dimensional point, and
$a$, $b$, and $c$ are the major, median, and minor semiaxes, satisfying
$a \ge b \ge c > 0$.  (The superscript $T$ means ``transpose,''
converting a row vector into a column vector.)  We characterize the {\it
shape} of the ellipsoid by the parameters,
\begin{equation}\label{param-def}
e  = \frac{\sqrt{a^2-c^2}}b,\quad
k  = \frac{\sqrt{b^2-c^2}}{\sqrt{a^2-c^2}},\quad
k' = \frac{\sqrt{a^2-b^2}}{\sqrt{a^2-c^2}}.
\end{equation}
Here $e$ measures how much the ellipsoid departs from a sphere, while
$k$ and $k'$ describe how close the ellipsoid is to being oblate ($k = 1$)
or prolate ($k = 0$); note that $k^2 + k'^2 = 1$.  The
semiaxes are related to these parameters by
\begin{equation}
[a,b,c] = b \bigl[ \sqrt{1 + e^2k'^2}, 1, \sqrt{1 - e^2k^2} \bigr].
\end{equation}
The case $c = 0$, where the ellipsoid becomes an elliptical disc, is
briefly discussed in Appendix \ref{app-billiards}.

A point on the ellipsoid can be written in terms of ellipsoidal
coordinates, the latitude, $\beta$, and the longitude, $\omega$, as
\begin{equation}\label{cart}
\mathbf R = \begin{bmatrix}
a \cos\omega \sqrt{k^2\cos^2\beta + k'^2}\\
b \cos\beta \sin\omega\\
c \sin\beta \sqrt{k^2 + k'^2\sin^2\omega}
\end{bmatrix}.
\end{equation}

\begin{figure}[tb]
\begin{center}
\includegraphics[scale=0.75]{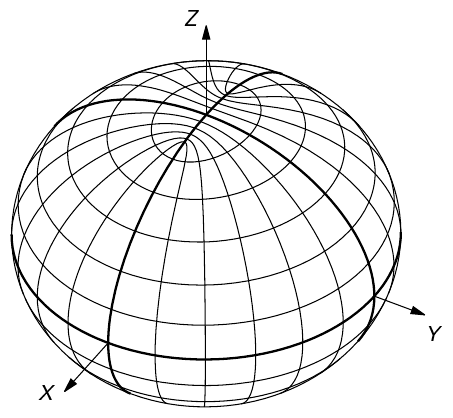}
\end{center}
\caption{\label{graticule}
The ellipsoidal grid showing lines of constant $\beta$ and $\omega$.
The grid spacing is $15^\circ$.  The heavy lines show the minor ($X = 0$
or $\cos\omega = 0$), median ($Y = 0$ or $\cos\beta \sin\omega = 0$),
and major ($Z = 0$ or $\sin\beta = 0$) principal ellipses of the
ellipsoid.  The parameters of the ellipsoid are $a = 1.01$, $b = 1$, and
$c = 0.8$, and it is viewed in an orthographic projection looking at the
point with geodetic coordinates $\phi = 40^\circ$, $\lambda = 30^\circ$.}
\end{figure}%
Lines of constant $\beta$ and $\omega$ define a grid, illustrated in
Fig.~\ref{graticule}.  The grid lines of the ellipsoidal coordinates are
``lines of curvature'' on the ellipsoid, i.e., they are parallel to the
directions of principal curvature.  The coordinates are singular at the
{\it umbilics}, $\cos\beta = \sin\omega = 0$, where the principal
curvatures are equal (locally, the ellipsoid is spherical).  The grid
lines are also intersections of the ellipsoid with confocal systems of
hyperboloids of one and two sheets.  Finally, the lines of curvature are
geodesic ellipses and hyperbolas, where the foci are neighboring
umbilics.

In the limit $k\rightarrow 1$ (resp.~$k\rightarrow 0$), the
umbilics converge on the $Z$ (resp.~$X$) axis and an
oblate (resp.~prolate) ellipsoid is obtained with $\beta$
(resp.~$\omega$) becoming the standard parametric latitude and
$\omega$ (resp.~$\beta$) becoming the standard longitude.  The
sphere is a non-uniform limit, with the position of the umbilics
depending on $k$.

Define three vectors giving the ``East,'' ``North,'' and ``Up''
directions:
\begin{subequations}
\begin{align}
\mathbf E &= \partial \mathbf R / \partial \omega
=\begin{bmatrix}
-a \sin\omega\sqrt{k^2\cos^2\beta + k'^2}\\[.5ex]
b\cos\beta\cos\omega\\[.5ex]
\displaystyle
c\frac{k'^2\sin\beta\cos\omega\sin\omega}
{\sqrt{k^2 + k'^2\sin^2\omega}}
\end{bmatrix},\displaybreak[0]\\
\mathbf N &= \partial \mathbf R / \partial \beta
=\begin{bmatrix}
\displaystyle
-a \frac{k^2 \cos\beta\sin\beta\cos\omega}
{\sqrt{k^2\cos^2\beta + k'^2}}\\[2.2ex]
-b \sin\beta\sin\omega\\[.5ex]
c \cos\beta\sqrt{k^2 + k'^2\sin^2\omega}
\end{bmatrix},\displaybreak[0]\\
\mathbf U &= \half\nabla S(\mathbf R)
=\biggl[
\frac X{a^2},
\frac Y{b^2},
\frac Z{c^2}
\biggr]^T.
\end{align}
\end{subequations}

It is easy to verify that $\mathbf N \cdot \mathbf E = 0$, so that
$[\mathbf E, \mathbf N, \mathbf U]$ are mutually orthogonal.  As a
consequence, the element of distance $\d s$ for the ellipsoidal
coordinate system is given by
\begin{align}
\frac{\d s^2}{b^2} &=
\frac{\abs{\mathbf N}^2 \, \d \beta^2 +
\abs{\mathbf E}^2 \, \d \omega^2}{b^2}\notag\\
&=
(k^2 \cos^2\beta + k'^2 \sin^2\omega)\notag\\
&\quad{}\times
\biggl(\frac{1 - e^2 k^2 \cos^2\beta}{k'^2 + k^2\cos^2\beta} \,\d \beta^2
+
\frac{1 + e^2 k'^2 \sin^2\omega}{k^2 + k'^2\sin^2\omega}
\,\d \omega^2\biggr).
\label{ds-eq}
\end{align}

Furthermore, the direction of a geodesic is
\begin{equation}\label{v1}
\mathbf V = \hat{\mathbf E}\sin\alpha + \hat{\mathbf N}\cos\alpha,
\end{equation}
where $\alpha$ is the azimuth of the geodesic measured clockwise from a
line of constant $\omega$ and the hat symbol denotes a unit vector.
At the pole of an oblate ellipsoid, we take the limit
$\cos\beta \rightarrow 0+$, to give
\begin{subequations}
\begin{align}
\hat{\mathbf E} &= [-\sin\omega, \cos\omega, 0]^T,\displaybreak[0]\\
\hat{\mathbf N} &= \sin\beta[-\cos\omega, -\sin\omega, 0]^T.
\end{align}
\end{subequations}
Similarly, at the pole of a prolate ellipsoid, we take the limit
$\sin\omega \rightarrow 0+$, to give
\begin{subequations}
\begin{align}
\hat{\mathbf E} &= \cos\omega[0, \cos\beta, \sin\beta]^T,\displaybreak[0]\\
\hat{\mathbf N} &= [0, -\sin\beta, \cos\beta]^T.
\end{align}
\end{subequations}

At an umbilic on a general ellipsoid, $\cos\beta \rightarrow 0$ and
$\sin\omega \rightarrow 0$, we have $\abs{\mathbf E} = \abs{\mathbf N} =
0$ so that $\hat{\mathbf E}$ and $\hat{\mathbf N}$ become ill-defined.
In this case, we use the conventional geodetic definitions of
$\hat{\mathbf E}$ and $\hat{\mathbf N}$,
\begin{subequations}
\begin{align}
\hat{\mathbf U} &= [ck' \cos\omega, 0, ak \sin\beta]^T/b,\displaybreak[0]\\
\hat{\mathbf E} &= [0, \cos\omega, 0]^T,\displaybreak[0]\\
\hat{\mathbf N} &= \hat{\mathbf U} \times \hat{\mathbf E}.
\end{align}
\end{subequations}
For geodesics that intersect an umbilic, we have
\begin{equation}\label{alpha-umb}
\tan\alpha = \pm \frac{k'}k \frac{\sin\omega}{\cos\beta};
\end{equation}
this follows from setting $\gamma = 0$ in Eq.~(\ref{gamma-def}), given
below.  Expanding $\mathbf R$ about an umbilic to second order in
$\cos\beta$ and $\sin\omega$, we find
\begin{equation}\label{v2}
\mathbf V = -\sin\beta(\hat{\mathbf E}\sin(2\alpha)
-\hat{\mathbf N}\cos(2\alpha)),
\end{equation}
where we have chosen the sign in Eq.~(\ref{alpha-umb}) as $\pm1 =
-\sin\beta\cos\omega$ to yield the normal convention that $\alpha$
measures angles clockwise.

The torus $(\omega, \beta) \in [-\pi,\pi] \times [-\pi,\pi]$ covers
the ellipsoid twice.  To facilitate passing to the limit of an
oblate ellipsoid, we may regard  $[-\pi,\pi]
\times [-\half\pi,\half\pi]$ as the principal sheet and insert branch cuts at
$\beta=\pm\half\pi$.  The rule for switching sheets is
\begin{equation}
\omega \rightarrow -\omega,\quad
\beta \rightarrow \pi-\beta,\quad
\alpha \rightarrow \pi+\alpha.
\end{equation}

Other coordinate systems are frequently used for an ellipsoid:
geodetic, parametric, and geocentric.  Conversions between the various
coordinate systems are considered in Appendix \ref{app-conversion}.

\section{Qualitative behavior}\label{qual-sec}

\begin{figure*}[tb!]
\begin{center}
\includegraphics[scale=0.75]{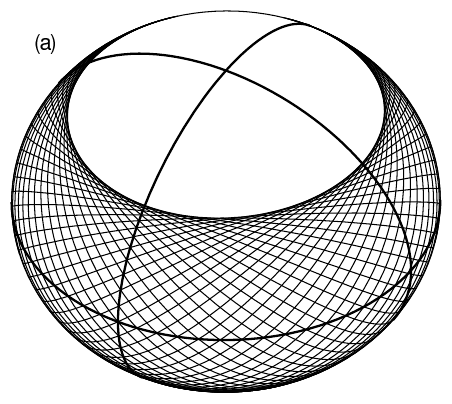}\hspace{20em}
\includegraphics[scale=0.75]{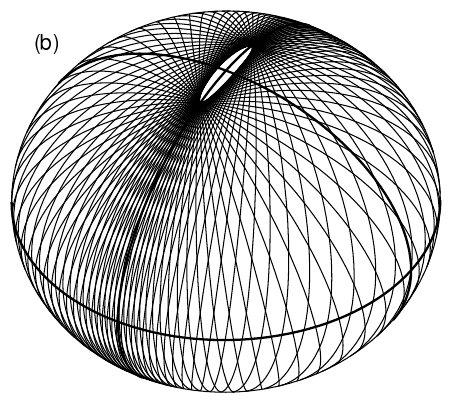}\\[-16ex]
\includegraphics[scale=0.75]{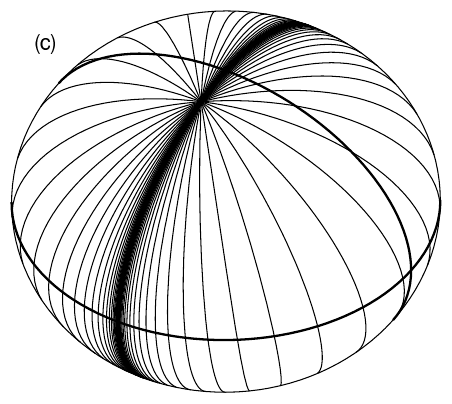}\\[-16ex]
\includegraphics[scale=0.75]{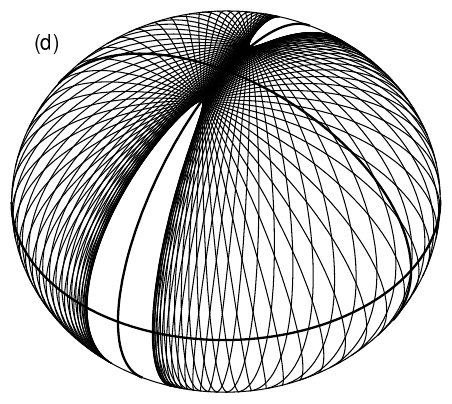}\hspace{20em}
\includegraphics[scale=0.75]{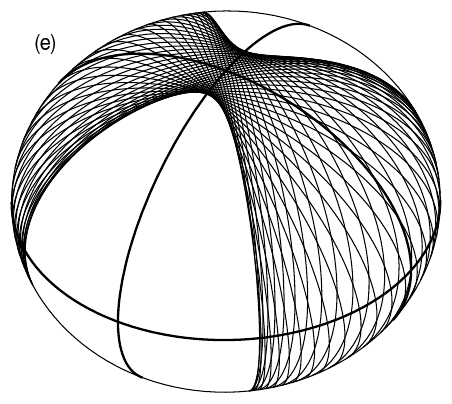}
\end{center}
\caption{\label{sample-fig}
Samples of geodesics on an ellipsoid with the same parameters and viewpoint
as Fig.~\ref{graticule}.  The parameters of the geodesics are given in
Table \ref{sample-tab}.  These figures are adapted from figures
that the author contributed to Wikipedia \citep{wikigeod}.}
\end{figure*}%
\begin{table}[tb]
\caption{\label{sample-tab}
Parameters for the sample geodesics shown in Fig.~\ref{sample-fig}.  The
starting points are given by $\beta_1$, $\omega_1$, and $\alpha_1$.  The
corresponding value of $\gamma$ is given by Eq.~(\ref{gamma-def}).  The
initial conditions are such that the starting points lie on the median
principal ellipse $Y = 0$ and the initial direction is $\mathbf V_1 =
[0,1,0]^T$.  The geodesics are followed a distance $\pm s_{12}$ in each
direction.  The ratio $r = p/q$ indicates that, over the full distance
$2s_{12}$, the geodesic executes $p$ complete oscillations/circuits in
$\beta$ and $q$ circuits/oscillations in $\omega$.}
\begin{tabular}{@{\extracolsep{0.4em}}
  c >{$}c<{$} >{$}c<{$} >{$}c<{$} >{$}c<{$} >{$}c<{$} >{$}c<{$}}
    \hline\hline\noalign{\smallskip}
& \beta_1\,(^\circ)
& \omega_1\,(^\circ)
& \alpha_1\,(^\circ)
& \gamma
& s_{12}
& r \\
\noalign{\smallskip}\hline\noalign{\smallskip}
(a) & 42.70330 &  0       &  90 &  0.51148 & 162.80637 & 61/52 \\
(b) & 87.52250 &  0       &  90 &  0.00177 & 247.24408 & 87/85 \\
(c) & 90       &  0       & 135 &  0       & 142.63587 & 50/50 \\
(d) & 90       & 10.15216 & 180 & -0.00164 & 252.96477 & 89/87 \\
(e) & 90       & 39.25531 & 180 & -0.02117 & 156.05191 & 55/53 \\
\noalign{\smallskip}
\hline\hline
\end{tabular}
\end{table}%
Let us illustrate the qualitative properties of geodesics; these are
readily found from the form of Jacobi's solution and are described in
the textbooks listed in the introduction.  On a given geodesic,
\begin{equation}\label{gamma-def}
\gamma = k^2\cos^2\beta\sin^2\alpha-k'^2\sin^2\omega\cos^2\alpha
\end{equation}
is a constant.  This is a generalization of the familiar Clairaut
constant, which characterizes geodesics on a biaxial ellipsoid.
Figure \ref{sample-fig} shows samples of geodesics on an ellipsoid with
$a = 1.01$, $b = 1$, and $c = 0.8$ (the same parameters as
Fig.~\ref{graticule}); the values of $\gamma$ for these geodesics are
given in Table~\ref{sample-tab}.  Ignoring for now
Fig.~\ref{sample-fig}(c), we see that, depending on whether $\gamma$ is
positive or negative, either $\omega$ or $\beta$ is a ``rotating''
coordinate (increasing or decreasing without limit), and,
correspondingly, $\beta$ or $\omega$ is a ``librating'' coordinate
(oscillating about a fixed value).  We label these two cases
``circumpolar'', $\gamma > 0$, Figs.~\ref{sample-fig}(a,~b), and
``transpolar'', $\gamma < 0$, Figs.~\ref{sample-fig}(d,~e).

The circumpolar geodesics are similar to the geodesics on an oblate
ellipsoid and a limiting case of such geodesics is the major
principal ellipse defined by $Z = 0$.  Likewise, the transpolar
geodesics mimic the geodesics on a prolate ellipsoid,
and a limiting case of such geodesics is the minor principal ellipse
defined by $X = 0$.  The transition between these two classes of
geodesics is shown in Fig.~\ref{sample-fig}(c), where $\gamma = 0$.  In
this case, the geodesic---an {\it umbilical geodesic}---repeatedly
crosses two opposite umbilics; following the geodesic in either
direction, it eventually lies on the median principal ellipse $Y = 0$.

For a biaxial ellipsoid, the equator and all the meridians are simple
(not self-intersecting) closed geodesics.  On the other hand, for a
triaxial ellipsoid, provided it is not too eccentric, there are only
three simple closed geodesics, namely the 3 principal ellipses.  (For
sufficiently eccentric ellipsoids, $e^2k^{3/2} > 3/4$, other simple
closed geodesics that oscillate about the major principal ellipse are
possible.)  The
major and minor ellipses are {\it stable}; if they are perturbed, the
resulting geodesic oscillates about the original ellipse.  However, the
median ellipse is {\it unstable}; if the geodesic is perturbed, it
swings away from the $Y = 0$ plane before returning to the original
ellipse, but now traveling along it in the opposite direction.  The
stability of closed geodesics is treated in Appendix \ref{closed-stab}.

For almost all $\gamma \ne 0$, a geodesic covers the area bounded by the
limiting lines of curvature.  The examples of geodesics in
Figs.~\ref{sample-fig}(a,~b,~d,~e) are exceptional in that
they are closed.  For the corresponding values of $\gamma \ne 0$ listed
in Table \ref{sample-tab}, the geodesics are closed in the same way
regardless of the initial conditions.  For a particular such $\gamma$,
the union of the closed geodesics is area filling; this is an example of
Poncelet's porism.  Umbilical geodesics, exemplified by
Fig.~\ref{sample-fig}(c), are not area filling, but here again, the
union of all such geodesics is, covering the entire ellipsoid.

\section{Jacobi's solution}

Here we summarize the solution of the geodesic problem
following \citet[\S\S583--585]{darboux94}; a comparable treatment is given
by \citet[\S\S3.5.4--3.5.6]{klingenberg82}.  The expression for $\d
s^2$ Eq.~(\ref{ds-eq}) fulfills the condition of a ``Liouville
surface,'' with metric given by Darboux's Eq.~(23),
\begin{equation}
\d s^2 = \bigl(U - V\bigr)
\bigl(U_1^2 \,\d u^2 + V_1^2 \,\d v^2\bigr),
\end{equation}
where $U$ and $U_1$ are functions of $u$ and $V$ and $V_1$ are functions
of $v$.  Identifying
\begin{subequations}
\begin{align}
(u, v) &= (\beta, \omega), \displaybreak[0]\\
(U, V) &= (k^2\cos^2\beta, -k'^2\sin^2\omega), \displaybreak[0]\\
(U_1^2, V_1^2) &= \biggl(
\frac{1 - e^2 k^2 \cos^2\beta}{k'^2 + k^2\cos^2\beta},
\frac{1 + e^2 k'^2 \sin^2\omega}{k^2 + k'^2\sin^2\omega}
\biggr),
\end{align}
\end{subequations}
the course of the geodesic is given by Darboux's Eq.~(28),
\begin{subequations}\label{jacobi-comb}
\begin{align}\label{course-eq}
\delta &= \int
\frac{\sqrt{1-e^2k^2\cos^2\beta}}
{\sqrt{k'^2+k^2\cos^2\beta}\sqrt{k^2\cos^2\beta-\gamma}}\,\d\beta\notag\\
&\qquad{}\mp\int
\frac{\sqrt{1+e^2k'^2\sin^2\omega}}
{\sqrt{k^2+k'^2\sin^2\omega}\sqrt{k'^2\sin^2\omega+\gamma}}\,\d\omega,
\end{align}
and the distance $s$ along the geodesic is given by Darboux's
Eq.~($33'$),
\begin{align}\label{dist-eq}
\frac{s+s_1}b &= \int
\frac{k^2\cos^2\beta \sqrt{1-e^2k^2\cos^2\beta}}
{\sqrt{k'^2+k^2\cos^2\beta}\sqrt{k^2\cos^2\beta-\gamma}}\,\d\beta\notag\\
&\quad{}\pm\int
\frac{k'^2\sin^2\omega \sqrt{1+e^2k'^2\sin^2\omega}}
{\sqrt{k^2+k'^2\sin^2\omega}\sqrt{k'^2\sin^2\omega+\gamma}}\,\d\omega.
\end{align}
\end{subequations}
Here $\delta$ and $s_1$ are constants given by the initial conditions.
Except at umbilics, the direction of the line is determined by the
constant $\gamma$, defined in Eq.~(\ref{gamma-def}) and given by
Darboux's Eq.~(30).  At umbilics, $\gamma$ vanishes, and the direction
is given by $\delta$.

The integrals in Eqs.~(\ref{jacobi-comb}) are related to one another.
It is therefore convenient to define
\begin{subequations} \label{comb-func}
\begin{align}
f(\phi;\kappa,\epsilon,\mu) &= \int_0
\frac{\sqrt{1-\epsilon\kappa\cos^2\phi}}
{\sqrt{\kappa'+\kappa\cos^2\phi}\sqrt{\kappa\cos^2\phi+\mu}}\,\d\phi,
\label{jacobi-func}\displaybreak[0]\\
g(\phi;\kappa,\epsilon,\mu) &= \int_0
\frac{\kappa\cos^2\phi\sqrt{1-\epsilon\kappa\cos^2\phi}}
{\sqrt{\kappa'+\kappa\cos^2\phi}\sqrt{\kappa\cos^2\phi+\mu}}\,\d\phi,
\label{dist-func}
\end{align}
\end{subequations}
where $\kappa \in [0,1]$, $\kappa' = 1-\kappa$, $\mu \in
[-\kappa; \kappa']$, $\epsilon \in (-\infty, 1/\kappa)$.  Equations
(\ref{jacobi-comb}) can be written as
\begin{subequations}\label{jacobi-ds}
\begin{align}
\delta &= f(\beta; k^2, e^2, -\gamma) \notag\\&\qquad{}
\mp f(\omega-\half\pi; k'^2, -e^2, \gamma),\label{jacobi}\displaybreak[0]\\
(s+s_1)/b &= g(\beta; k^2, e^2, -\gamma) \notag\\&\qquad{}
\pm g(\omega-\half\pi; k'^2, -e^2, \gamma).\label{ds}
\end{align}
\end{subequations}
At present, we leave the signs of the square roots in the integrals
unspecified.  However, the presence of $\mp$ and $\pm$ in these
equations indicates that while progressing along a geodesic, the two
terms in Eq.~(\ref{jacobi}) cancel while those in Eq.~(\ref{ds})
combine.  In the following, we drop the parametric arguments for the $f$
and $g$ functions; $[\kappa, \epsilon, \mu] = [k^2, e^2, -\gamma]$ are
implied for functions of $\beta$, and $[k'^2, -e^2, \gamma]$ for
functions of $\omega$.

\begin{figure*}[tb]
\begin{center}
\includegraphics[scale=0.75]{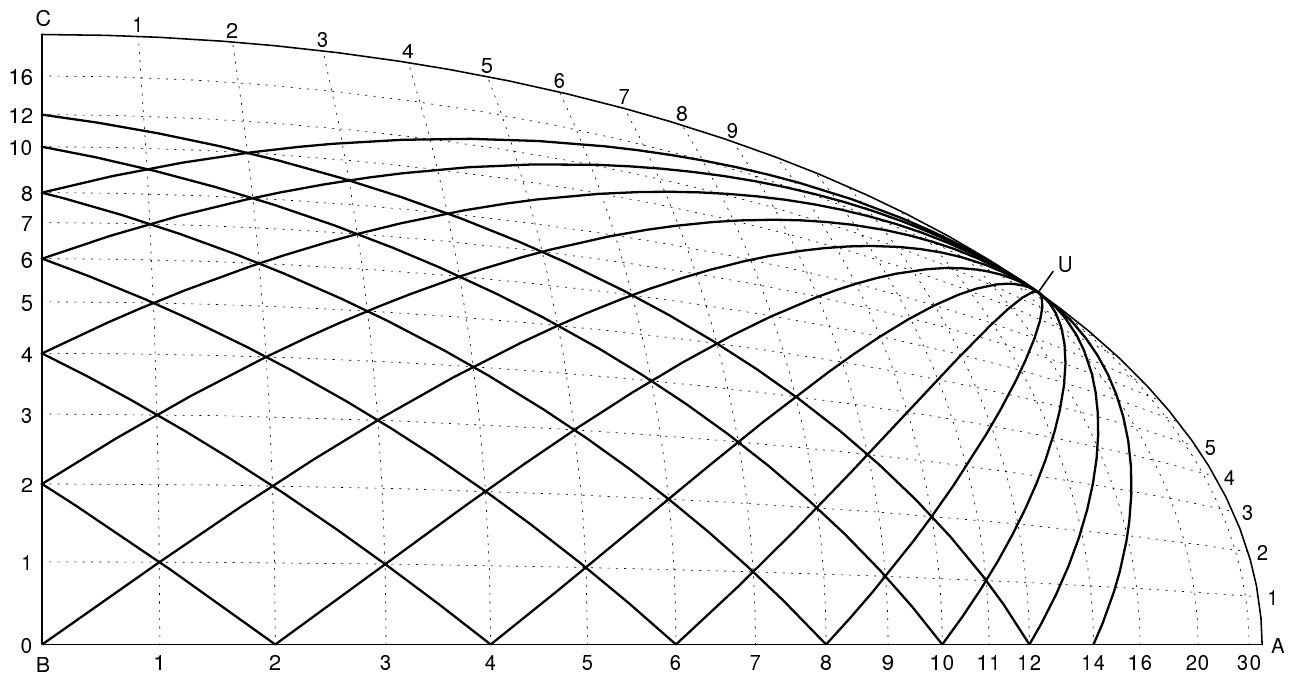}
\end{center}
\caption{\label{cayley-fig}
The graphical method for plotting umbilical geodesics given
by \citet{cayley72}.  The semiaxes of the ellipsoid are $a
= \sqrt{1000}$, $b = \sqrt{500}$, $c = \sqrt{250}$, and it is viewed
here in an orthogonal projection along the $Y$ axis.  The dotted lines
are lines of $\beta = \beta^{(f)}_n$ (going bottom to top) or $\omega
= \omega^{(f)}_n$ (going left to right), as defined in the text, with
the separation constant given by $\Delta^{(f)} = 1/\sqrt{160}$ (matching
Cayley's choice).  In this figure, the lines of constant $\beta$ and
$\omega$ are labeled with the corresponding values of $n$ (these values
are given in Table \ref{cayley-tab}).  The geodesics, shown as heavy
lines, connect the vertices of the resulting mesh, and all converge on
the umbilic labeled $U$, or on the neighboring one (not shown in the
figure).}
\end{figure*}%
\begin{figure*}[tb]
\begin{center}
\includegraphics[scale=0.75]{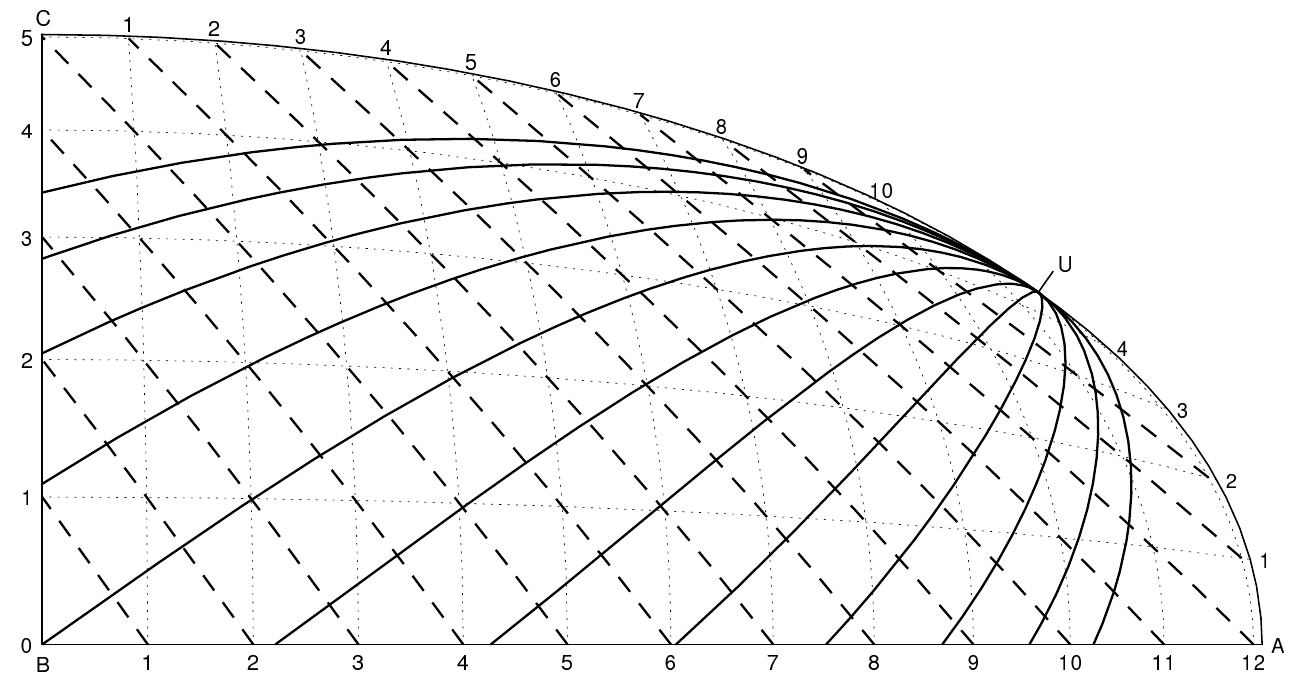}
\end{center}
\caption{\label{cayleydist-fig}
Marking the distance along geodesics.  The heavy lines are those
geodesics in Fig.~\ref{cayley-fig} that converge on the umbilic
$U$.  As in that figure, the dotted lines are lines of constant $\beta$
or $\omega$; however, in this case, the values are given by equal
increments of $\Delta^{(g)} = 1/10$ in the distance functions (these
values are given in Table \ref{cayley-tab}).  The dashed lines connect
the vertices of the mesh, and these mark off distance intervals
of $b/10$ along the geodesics.}
\end{figure*}%
\begin{table}[tbp]
\caption{\label{cayley-tab}
Values of $\beta^{(f)}_n$, $\omega^{(f)}_n$, $\beta^{(g)}_n$, and
$\omega^{(g)}_n$ used for the coordinate meshes in
Figs.~\ref{cayley-fig} and \ref{cayleydist-fig}.  For the coordinates in
the 2nd and 3rd columns, the values of the $f$ functions are multiples
of $\Delta^{(f)} = 1/\sqrt{160}$, and the coordinates are used in
Fig.~\ref{cayley-fig}.  For those in the 4th and 5th columns, the values
of the $g$ functions are multiples of $\Delta^{(g)} = 1/10$, and the
coordinates are used in Fig.~\ref{cayleydist-fig}. }
\newcommand{\0}{\hphantom{0}}
\begin{tabular}{@{\extracolsep{1.4em}}
  >{$}r<{$} >{$}l<{$} >{$}l<{$} >{$}l<{$} >{$}l<{$}}
    \hline\hline\noalign{\smallskip}
n
& \beta^{(f)}_n\,(^\circ)
& \omega^{(f)}_n\,(^\circ)
& \beta^{(g)}_n\,(^\circ)
& \omega^{(g)}_n\,(^\circ) \\
\noalign{\smallskip}\hline\noalign{\smallskip}
  0 &\0  0     &\0  90     &\0  0     &\0  90     \\
  1 &\0  7.789 &\0  95.538 &   13.993 &\0  94.967 \\
  2 &   15.265 &   101.015 &   27.852 &\0  99.966 \\
  3 &   22.205 &   106.377 &   41.915 &   105.029 \\
  4 &   28.511 &   111.571 &   57.515 &   110.195 \\
  5 &   34.175 &   116.557 &   84.901 &   115.507 \\
  6 &   39.237 &   121.301 &          &   121.023 \\
  7 &   43.758 &   125.782 &          &   126.818 \\
  8 &   47.803 &   129.985 &          &   133.004 \\
  9 &          &   133.906 &          &   139.762 \\
 10 &   54.696 &   137.547 &          &   147.434 \\
 11 &          &   140.916 &          &   156.870 \\
 12 &   60.309 &   144.024 &          &   173.205 \\
 14 &          &   149.518 \\
 16 &   68.769 &   154.156 \\
 20 &          &   161.362 \\
 30 &          &   171.645 \\
\noalign{\smallskip}
\hline\hline
\end{tabular}
\end{table}%
The structure of Eqs.~(\ref{jacobi}) allows geodesics to be traced by a
simple construction given by \citet{cayley72}, who considered umbilical
geodesics, $\gamma = 0$, on an ellipsoid with $a:b:c
= \sqrt2:1:1/\sqrt2$.  Find the values of $\beta$ (resp.~$\omega$), such
that $f(\beta^{(f)}_n) = n\Delta^{(f)}$
(resp.~$f({\omega^{(f)}_n-\half\pi}) = n\Delta^{(f)}$).  Now draw the
grid lines $\beta = \beta^{(f)}_n$ and $\omega = \omega^{(f)}_n$,
forming a mesh on the ellipsoid.  Two families of geodesics can be
traced through the mesh by connecting opposite corners of each cell.
This follows from Eq.~(\ref{jacobi}), and the method only ``works''
because of the separation of variables in the solution.  The result is
shown in Fig.~\ref{cayley-fig}, where we reproduce the case examined
by \citet[Plate II, following p.~130]{cayley72}.  (Besides some
understandable errors arising from the low-order methods he used to
evaluate and invert the integrals, Cayley made some mistakes connecting
the vertices of the mesh.)

The same construction can be used to mark off distances along the
geodesics using Eq.~(\ref{ds}).  In this case, we construct a
mesh defined by $g(\beta^{(g)}_n) = n\Delta^{(g)}$ and
$g({\omega^{(g)}_n-\half\pi}) = n\Delta^{(g)}$ as shown in
Fig.~\ref{cayleydist-fig}.  We see that the solution of the direct
geodesic problem essentially reduces to ``tabulating'' four one-dimensional
integrals.

\section{Properties of general geodesics}\label{general-sec}

The course of a geodesic is determined by Eq.~(\ref{jacobi}) and we are
immediately confronted with the problem that, for $\gamma \neq 0$, one
of the integrands in Eq.~(\ref{course-eq}) is singular and that, if
$\gamma = 0$, the integrals themselves are singular.  In this section,
we address the circumpolar and transpolar cases where $\gamma \neq 0$.

We deal with these cases together by denoting the rotating and librating
coordinates as $\theta$ and $\phi$, respectively.  We include the offset
of $\half\pi$ in relations involving $\omega$.  Let us also define
$\tau$ as the azimuth measured from a line of constant $\theta$.
Because $\phi$ and $\tau$ are librating coordinates, it's useful to
introduce constants $S_\phi = \sign(\cos\phi)$ and $S_\tau
= \sign(\sin\tau)$ to specify the values about which $\phi$ and $\tau$
oscillate.

For the rotating coordinate $\theta$, we fold in the
direction of the geodesic so that $\theta$ increases in the forward
direction.  Thus, for circumpolar geodesics, we have
\begin{subequations}
\begin{equation}
\phi = \beta, \quad \tau = \alpha, \quad \theta = S_\tau(\omega-\half\pi),
\end{equation}
while for transpolar geodesics,
\begin{equation}
\phi = \omega-\half\pi, \quad \tau = \half\pi-\alpha, \quad
\theta = S_\tau\beta.
\end{equation}
\end{subequations}
We define
\begin{subequations} \label{fgtheta}
\begin{align}
f_\theta(\theta) &= \int_0
\frac{\sqrt{1-\epsilon\kappa\cos^2\theta}}
{\sqrt{\kappa'+\kappa\cos^2\theta}
\sqrt{\kappa\cos^2\theta+\abs\mu}}\,\d\theta,
\label{ftheta}\displaybreak[0]\\
g_\theta(\theta) &= \int_0
\frac{\kappa\cos^2\theta\sqrt{1-\epsilon\kappa\cos^2\theta}}
{\sqrt{\kappa'+\kappa\cos^2\theta}
\sqrt{\kappa\cos^2\theta+\abs\mu}}\,\d\theta,
\label{gtheta}
\end{align}
\end{subequations}
replacing $\phi$ by $\theta$ in Eqs.~(\ref{comb-func}) and stipulating
that positive square roots are to be taken in the integrands.  We
have replaced $\mu$ by $\abs\mu$, consistent with the requirement that
$0 < \mu \le \kappa'$ for the $\theta$ integrals.

In the $\phi$ integrals in Eqs.~(\ref{comb-func}), we have
$-\kappa \le \mu < 0$ leading to a weak (square-root) singularity in the
integrand at $\cos\phi = \sqrt{\abs\mu/\kappa}$.  This singularity can
be removed by changing the variable of integration to $\psi$ defined by
\begin{subequations}
\begin{align}
\sin\phi &= \sin\psi \sqrt{1 - \abs\mu/\kappa}, \displaybreak[0]\\
\frac{\d\phi}{\d\psi} &= \frac{\cos\psi}{\cos\phi}
\sqrt{1 - \abs\mu/\kappa}.
\end{align}
\end{subequations}
Note that $\d\phi/\d\psi$ changes sign at the vertices of the geodesic,
where $\cos\psi = 0$ or, equivalently, where $\cos\tau = 0$.  It is
convenient to write $\psi$ in terms of $\tau$ and $\theta$ by rewriting
Eq.~(\ref{gamma-def}) as
\begin{equation}
-\mu = \kappa\cos^2\phi\sin^2\tau - \kappa'\cos^2\theta\cos^2\tau.
\end{equation}
The interconversions of $\psi$, $\phi$, and $\tau$ become
\begin{subequations}
\begin{align}
\psi &= \atanx{\sqrt\kappa \sin\phi}
{S_\phi \cos\tau \sqrt{\kappa\cos^2\phi + \kappa' \cos^2\theta}},
\displaybreak[0]\\
\phi &= \atanx{\sqrt{\kappa-\abs\mu} \sin\psi}
{S_\phi \sqrt{\kappa \cos^2\psi + \abs\mu \sin^2\psi}},
\displaybreak[0]\\
\tau &= \atanx{S_\tau\sqrt{\kappa' \cos^2\theta + \abs\mu}}
{S_\phi\sqrt{\kappa-\abs\mu} \cos\psi}.
\end{align}
\end{subequations}
The heavy ratio line in the argument to the arctangent indicates that
the quadrant of the function is given by the signs of the numerator and
denominator separately.  This ensures that $\psi$ increases in the
forward direction along a geodesic.

The functions $f(\phi)$ and $g(\phi)$ are replaced by
\begin{subequations}\label{fgstar}
\begin{align}
f_\psi(\psi) &=
\int_0
\frac{\sqrt{1-\epsilon(\kappa\cos^2\psi + \abs\mu\sin^2\psi)}}
{\sqrt{\kappa' + \kappa\cos^2\psi + \abs\mu\sin^2\psi}} \notag\\
&\qquad\times
\frac{\d\psi}{ \sqrt{\kappa\cos^2\psi + \abs\mu\sin^2\psi}},
\label{fstar}\displaybreak[0]\\
g_\psi(\psi) &=
\int_0
\frac{\sqrt{1-\epsilon(\kappa\cos^2\psi + \abs\mu\sin^2\psi)}}
{\sqrt{\kappa' + \kappa\cos^2\psi + \abs\mu\sin^2\psi}} \notag\\
&\qquad\times
 \sqrt{\kappa\cos^2\psi + \abs\mu\sin^2\psi}\,\d\psi;
\label{gstar}
\end{align}
\end{subequations}
positive square roots should be taken in these integrals.  The variable
$\psi$ (which, like $\theta$, is a rotating coordinate) plays the same
role as the arc length on the auxiliary sphere in Bessel's solution
of the geodesic problem on a biaxial ellipsoid; it allows the geodesic
to be tracked through its vertices (points of extreme latitude) and to
be followed indefinitely.

Now, the geodesic is given by
\begin{subequations}\label{comb-mod}
\begin{align}
\delta &= f_\psi(\psi) - f_\theta(\theta),
\label{jacobi-mod}\displaybreak[0]\\
(s+s_1)/b &= g_\psi(\psi) + g_\theta(\theta).\label{ds-mod}
\end{align}
\end{subequations}
The omitted parameters for $f_\psi$ and $g_\psi$ are
$[\kappa, \epsilon, \mu] = [k^2, e^2, -\gamma]$ for circumpolar
geodesics and $[k'^2, -e^2, \gamma]$ for transpolar geodesics;
conversely, the omitted parameters for $f_\theta$ and $g_\theta$ are
$[k^2, e^2, -\gamma]$ for transpolar geodesics and $[k'^2,
-e^2, \gamma]$ for circumpolar geodesics.

The integrands in the definitions of these functions are analytic,
positive, even, and periodic with period $\pi$; so the functions are
analytic, odd, increasing functions, consisting of a secular term with a
superimposed periodic ripple.  (The integrand for $g_\theta$ vanishes
for $\cos\theta = 0$; so $g_\theta$ is merely non-decreasing.)  We
discuss the numerical evaluation of these integrals in
Sec.~\ref{int-eval}.

It is clear that we have a complete solution to the direct problem
for $\gamma \ne 0$.  The initial conditions, $s = 0$, $\beta_1$,
$\omega_1$, $\alpha_1$, allow the constants $\delta$ and $s_1$ in
Eqs.~(\ref{comb-mod}) to be determined.  For a
given distance from the initial point $s = s_{12} = s_2 - s_1$, these
equations have a unique solution for the endpoint $\theta_2$ and
$\phi_2$, which allows the coordinates and azimuth at point 2 to be
found.

\section{Properties of umbilical geodesics}\label{umb-sec}

Umbilical geodesics are characterized by $\gamma = 0$.  In this case,
the two integrals in Eq.~(\ref{jacobi-mod}) have logarithmic
singularities at $\cos\theta = 0$ and $\cos\psi = 0$.  These
singularities cancel at umbilics where both $\cos\theta$ and $\cos\psi$
vanish.  A geodesic leaving a particular umbilic in a specified
direction must arrive at the opposite umbilic with a well-defined
azimuth.  We use a connection relation to determine the azimuth on
leaving that umbilic.  The process can be repeated to follow a geodesic
through multiple passages of the umbilics.

We treat the umbilical geodesics as the limiting case of circumpolar
geodesics, i.e., $\gamma \rightarrow 0+$.  (Treating the other case,
$\gamma \rightarrow 0-$, follows a comparable procedure.)  Thus, $\psi$
and $\theta$ are related to $\beta$ and $\omega$, respectively.  With
$\gamma = 0$, both $(f_\psi, g_\psi)$ and $(f_\theta, g_\theta)$ have
the same functional forms, with, for example, Eqs.~(\ref{fgstar})
becoming
\begin{subequations}\label{fgumb}
\begin{align}
f_\psi(\psi;\mu=0) &= \int_0
\frac{\sqrt{1-\epsilon\kappa\cos^2\psi}}
{\sqrt{\kappa'+\kappa\cos^2\psi}\sqrt\kappa\cos\psi}\,\d\psi,
\label{fumb}\displaybreak[0]\\
g_\psi(\psi;\mu=0) &= \int_0
\frac{\sqrt\kappa\cos\psi\sqrt{1-\epsilon\kappa\cos^2\psi}}
{\sqrt{\kappa'+\kappa\cos^2\psi}}\,\d\psi.
\label{gumb}
\end{align}
\end{subequations}
We have substituted $\sqrt{\cos^2\psi} = \cos\psi$, because,
within each geodesic segment, we will take $\abs{\psi}\le \half\pi$, so
that $\cos\psi \ge 0$.

The integrals Eqs.~(\ref{fgumb}) can be given in closed form in the
spherical limit, $\epsilon = 0$, giving
\begin{subequations}\label{fgk00}
\begin{align}
\label{fk00}
f_\psi(\psi; \epsilon=0, \mu=0) &= \int_0
\frac1
{\sqrt{\kappa'+\kappa\cos^2\psi}\sqrt{\kappa}\cos\psi}\,\d\psi \notag\\
&=\frac{\sinh^{-1}(\sqrt{\kappa'}\tan\psi)}{\sqrt{\kappa\kappa'}},
\displaybreak[0]\\
\label{gk00}
g_\psi(\psi; \epsilon=0, \mu=0) &= \int_0
\frac{\sqrt\kappa\cos\psi}
{\sqrt{\kappa'+\kappa\cos^2\psi}}\,\d\psi \notag\\
&=\tan^{-1}\frac{\sqrt{\kappa}\sin\psi}{\sqrt{\kappa' + \kappa\cos^2\psi}}.
\end{align}
\end{subequations}
We write
\begin{subequations}\label{umbfdf}
\begin{align}
f_\psi(\psi;\mu = 0) &=
f_\psi(\psi;\epsilon = 0, \mu = 0)-
\Delta f_\psi(\psi;\mu = 0),\label{umbf}\displaybreak[0]\\
\Delta f_\psi(\psi;\mu = 0) &= \int_0
\frac{\epsilon\sqrt\kappa\cos\psi}
{\sqrt{\kappa'+\kappa\cos^2\psi}
\bigl(1+\sqrt{1-\epsilon\kappa\cos^2\psi}\bigr)}\,\d\psi.\label{delta-f0}
\end{align}
\end{subequations}
We could also express $g_\psi(\psi;\mu=0)$ as the sum of
$g_\psi(\psi; \allowbreak {\epsilon=0}, {\mu=0})$ and a correction, but
this is not necessary for its accurate evaluation.  The integrand in
Eq.~(\ref{delta-f0}) is free of singularities, so Eq.~(\ref{fk00})
captures the full singular behavior of $f_\psi(\psi;{\mu = 0})$.  In the
limit $\psi \rightarrow \pm\half\pi$, we have
\begin{equation}\label{umb-sing}
f_\psi(\psi;\mu = 0) \rightarrow
\pm \frac{\log( 2 \sqrt{\kappa'} \sec\psi)}{\sqrt{\kappa\kappa'}}.
\end{equation}
For each of Eqs.~(\ref{fgumb})--(\ref{umb-sing}), we have a
corresponding equation substituting $\theta$ for $\psi$.  In the
functions of $\psi$, the implied parameters are $[\kappa, \epsilon] =
[k^2, e^2]$, and for the functions of $\theta$, the parameters are
$[k'^2, -e^2]$.

Label each geodesic segment from one umbilic to the next sequentially
with index $j$.  We use superscripts $\mp$ to label the start ($\theta
= \psi = -\half\pi$) and end ($\theta = \psi = \half\pi$) of a segment.
At either end of the $j$th segment, we have
\begin{align}
\delta_j &=
\pm\biggl(\frac{1}{kk'}\log\frac{k'\sec\psi_j^{\pm}}{k\sec\theta_j^{\pm}}
-
\Delta f_\psi(\half\pi) + \Delta f_\theta(\half\pi)\biggr)
\notag\\
&= \pm\frac{1}{kk'} \biggl(\log\abs{\tan\alpha_j^\pm} - \half\Delta\biggr),
\label{delta}
\end{align}
where
\begin{equation}
\Delta = 2kk' \bigl( \Delta f_\psi(\half\pi)
   - \Delta f_\theta(\half\pi)\bigr). \label{Delta}
\end{equation}
We have introduced $\alpha_j^\pm$ from Eq.~(\ref{alpha-umb}), which
gives
\begin{equation}
\frac{k'\sec\psi}{k\sec\theta} = \pm\tan\alpha.
\end{equation}
Equating $\delta_j$ at the two ends of the umbilical segment gives
\begin{equation}
\tan\alpha_j^- \tan\alpha_j^+ = \exp(\Delta),\label{alpha-conn}
\end{equation}
where $\alpha_j^\pm$ lie in the same quadrant.

To connect to the next segment, we need to jump over the umbilic.
Consider the point near the end of a geodesic segment at
$(\beta_j^+, \omega_j^+)$.  The point at the start of the next segment
at $(\beta_{j+1}^-, \omega_{j+1}^-)$ is on the diametrically opposite
side of the umbilic and is given by
\begin{equation}
\cos\beta_{j+1}^- = \pm \frac{k'}k \sin\omega_j^+,\qquad
\sin\omega_{j+1}^- = \mp \frac k{k'} \cos\beta_j^+,
\end{equation}
with the sign of $\cos\beta$ preserved.  The azimuths are related by
\begin{equation}
\tan\alpha_j^+ \tan\alpha_{j+1}^- = -1,\qquad
\alpha_{j+1}^- - \alpha_j^+ = \pm \half\pi,
\end{equation}
with the sign chosen to preserve the sign of $\sin\alpha$.

This allows us to express $\delta_{j+1}$, using the lower signs in
Eq.~(\ref{delta}), in terms of $\delta_j$, using the upper signs in
Eq.~(\ref{delta}), which gives
\begin{equation}
\delta_{j+1} = \delta_j + \frac\Delta{kk'}.
\end{equation}
In general, we obtain
\begin{subequations}\label{alpha-j}
\begin{align}
\delta_j &= \delta_0 + j \frac\Delta{kk'},\label{delta-j}\displaybreak[0]\\
\tan\alpha_j^- &= (-1)^j\exp(-j\Delta)\tan\alpha_0^-,
\label{alpha-jm}\displaybreak[0]\\
\tan\alpha_j^+ &= (-1)^j\exp\bigl((j+1)\Delta\bigr)/\tan\alpha_0^-.\label{alpha-jp}
\end{align}
\end{subequations}

The presence of the exponential terms in Eqs.~(\ref{alpha-j}) is
evidence of the instability of umbilical geodesics discussed in
Sec.~\ref{qual-sec}.  This property of umbilical geodesics was
discovered by \citet{hart49a}, who gives an alternative (but equivalent)
expression for $\Delta$,
\begin{equation}
\Delta = \int_0^{\pi/2} \frac{\sqrt{(a^2 - b^2)(b^2 - c^2)}}
{\sqrt{a^2\tan^2\phi + b^2}\sqrt{c^2\tan^2\phi + b^2}}\,
\d\phi.\label{Delta-hart}
\end{equation}

The integrands in Eqs.~(\ref{gumb}) and (\ref{delta-f0}) are analytic,
even, and periodic with period $2\pi$.  In addition, they are odd about
the point $\theta = \half\pi$ and positive for $\abs{\theta}
< \half\pi$.  Thus, the integrals are analytic, odd, periodic with
period $2\pi$, and increasing in the interval $\abs{\theta} < \half\pi$.
The distance between opposite umbilics is
\begin{equation}
s_0 = 2b\bigl(g_\psi(\half\pi; \mu = 0) + g_\theta(\half\pi; \mu = 0)\bigr),
\label{umb-dist}
\end{equation}
half the perimeter of the median principal ellipse.

When solving the direct problem, $\delta$ and $s_1$ are determined as
before, with the proviso that, if the initial point is an umbilic,
$\delta = \delta_0$ should be evaluated using Eq.~(\ref{delta}) and the
initial azimuth.  With a given $s_{12}$, determine $s_2 = s_1 + s_{12}$
and find the umbilical segment index with
\begin{equation}
j = \round{s_2/s_0},
\end{equation}
where $\round x$ is the nearest integer to $x$.  Then solve the geodesic
equations with $s = s_2 - js_0$ and $\delta = \delta_j$ found with
Eq.~(\ref{delta-j}).  If initial conditions are such that the geodesic
lies on the median principal ellipse, then $\delta = \pm\infty$
diverges.  In this case, the geodesic can be broken into segments of
lengths $2b g_\psi(\half\pi; {\mu = 0})$ and $2bg_\theta(\half\pi; {\mu
= 0})$, the distances between neighboring umbilics.

\section{Biaxial ellipsoids}

Geodesics on a biaxial ellipsoid are well understood.  However, it's
instructive to see how we can recover the biaxial solution from
Jacobi's.

For biaxial ellipsoids, we substitute $\kappa = 0$ in
Eqs.~(\ref{fgtheta}) to give
\begin{subequations}\label{fgtheta-bi}
\begin{align}
f_\theta(\theta; \kappa = 0) &= \frac\theta{\sqrt{\abs\gamma}},
\label{ftheta-bi}\displaybreak[0]\\
g_\theta(\theta; \kappa = 0) &= 0.
\label{gtheta-bi}
\end{align}
\end{subequations}
Similarly, we substitute $\kappa = 1$ in Eqs.~(\ref{fgstar}) to obtain
\begin{subequations}\label{fgstar-bi}
\begin{align}
f_\psi(\psi; \kappa = 1) &=
\int_0
\frac{\sqrt{1-\epsilon(\cos^2\psi + \abs\gamma\sin^2\psi)}}
{\cos^2\psi + \abs\gamma\sin^2\psi}\,\d\psi,
\label{fstar-bi}\displaybreak[0]\\
g_\psi(\psi; \kappa = 1) &=
\int_0
\sqrt{1-\epsilon(\cos^2\psi + \abs\gamma\sin^2\psi)}\,\d\psi.
\label{gstar-bi}
\end{align}
\end{subequations}
In both Eqs.~(\ref{fgtheta-bi}) and (\ref{fgstar-bi}), I have replaced
$\abs\mu$ by $\abs\gamma$ in order to facilitate their substitution into
Eqs.~(\ref{comb-mod}).
In the spherical limit, we carry out the integration to give
\begin{subequations}\label{fg10m}
\begin{align}
f_\psi(\psi; \kappa=1, \epsilon=0) &=
\int_0 \frac 1{\cos^2\psi + \abs\gamma\sin^2\psi}\,\d\psi\notag\\
&=\frac1{\sqrt{\abs\gamma}} \atanx{\sqrt{\abs\gamma}\sin\psi}{\cos\psi},
\label{f10m}
\displaybreak[0]\\
g_\psi(\psi; \kappa=1, \epsilon=0) &= \psi. \label{g10m}
\end{align}
\end{subequations}
(The arctangent function in Eq.~(\ref{f10m}) tracks the quadrant of
$\psi$ through multiple revolutions.)  Following the same procedure used
in the umbilical case, Eqs.~(\ref{umbfdf}), we write
\begin{subequations}\label{fdf-obl}
\begin{align}
f_\psi(\psi; \kappa = 1) &=
f_\psi(\psi; \kappa = 1, \epsilon = 0)
- \Delta f_\psi(\psi; \kappa = 1),\label{f-obl}
\displaybreak[0]\\
\Delta f_\psi(\psi; \kappa = 1) &=\int_0 \frac\epsilon
{1+\sqrt{1-\epsilon(\cos^2\psi\abs\gamma\sin^2\psi)}}\,\d\psi.\label{df-obl}
\end{align}
\end{subequations}
The geodesic equations Eqs.~(\ref{comb-mod}) become
\begin{subequations}\label{comb-bi}
\begin{align}
\theta + \delta\sin\tau_0 &= \atanx{\sin\tau_0\sin\psi}{\cos\psi} \notag\\
&\quad{}-
\int_0 \frac{\epsilon\sin\tau_0}
{1+\sqrt{(1-\epsilon) + \epsilon\cos^2\tau_0\sin^2\psi}}
,\label{jacobi-bi}\displaybreak[0]\\
s + s_1 &= b\int_0
\sqrt{(1-\epsilon) + \epsilon\cos^2\tau_0\sin^2\psi}\,\d\psi,\label{ds-bi}
\end{align}
\end{subequations}
where $\sin^2\tau_0 = \abs\gamma = \cos^2\phi\sin^2\tau$, $\epsilon =
e^2 = ({a^2 - c^2})/a^2$ for oblate ellipsoids, and $\epsilon = -e^2 =
({c^2 - a^2})/c^2$ for prolate ellipsoids.  It is readily seen that these
agree with the standard formulas for biaxial ellipsoids, e.g., Eqs.~(8)
and (7) of \citet{karney13}.

We treat meridional geodesics on a biaxial ellipsoid by the same
mechanisms used for umbilical geodesics in Sec.~\ref{umb-sec}.  In the
limit $\gamma \rightarrow 0$, Eq.~(\ref{f10m}) becomes
\begin{equation}
f_\psi(\psi; \kappa = 1, \epsilon = 0, \mu \rightarrow 0) =
\frac{\round{\psi/\pi}\pi}{\sqrt{\abs\gamma}}
+ \tan\psi.\label{f0-obl-exp}
\end{equation}
Equation (\ref{jacobi-mod}) gives the expected result that $\theta$ is
constant except on passage through a pole, $\cos\psi = 0$, where
$\theta$ increases by $\pi$.  Passages through the poles are at
intervals of $s_0 = 2bg_\psi({\half\pi; {\kappa = 1}, {\mu = 0}})$.  The
solution of the direct problem proceeds analogously to the umbilical
case.

When tackling the inverse problem, we will need to find meridional
conjugate points.  (This is needed to determine whether the shortest
geodesic for two points on opposite meridians follows the meridian.)
The conjugate points correspond to a change in $\theta$ by a multiple of
$\pi$, giving
\begin{equation}
f_\theta(\theta_2; \kappa = 0) -
f_\theta(\theta_1; \kappa = 0) = \frac{n\pi}{\sqrt{\abs\gamma}}.
\end{equation}
To cancel this singular change in $f_\theta$, $\psi$ must cross over $n$
poles.  Assuming that $\psi_1\in(-\half\pi,\half\pi)$, we have
$\psi_2\in\bigl(({n-\half})\pi,\allowbreak({n+\half})\pi\bigr)$.  To
find the value of $\psi$ within the allowed range, we balance the
second non-singular term in Eq.~(\ref{f0-obl-exp}) against $\Delta
f_\psi$, i.e., we solve
\begin{align}
\tan\psi_2 - \tan\psi_1
&= \Delta f_\psi(\psi_2; \kappa = 1, \mu = 0)\notag\\
&\qquad{}- \Delta f_\psi(\psi_1; \kappa = 1, \mu = 0),
\label{meridconj}
\end{align}
for $\psi_2$ (in the allowed range).  This is accomplished by solving
for $y = \tan\psi_2$ using Newton's method.

\section{Evaluating the integrals}\label{int-eval}

Jacobi's solution reduces the original geodesic problem, coupled
ordinary differential equations, to the evaluation of one-dimensional
integrals.  Jacobi identifies these as abelian integrals, but this does
not particularly help because there are no simple procedures for
computing them.

We are therefore left with numerical quadrature of some sort.  This is
exactly the approach taken by \citet{cayley72} to construct graphically
the paths of umbilical geodesics (as shown in Fig.~\ref{cayley-fig}).
More recently, \citet{baillard15} provides routines for the HP-41
calculator for solving the inverse geodesic problem by performing the
corresponding definite integrals for Jacobi's solution using
Gauss-Legendre quadrature.

In this work, I sought a method for evaluating the integrals that
allows the solution to be found with nearly full double-precision
accuracy.  We would also like to be able to compute the {\it
indefinite} integrals rapidly, so that points at arbitrary positions on
a geodesic can be found.  The method should lend itself to implementation
at higher precision with a corresponding increase in accuracy at a
modest cost.

The concept of indefinite integration in numerical applications was
introduced by \citet{clenshaw60}, who showed that, having computed a
definite integral over the range $[a, b]$, it is possible with ``little
extra complication'' to determine the integral over any interior
interval.  Their method naturally extends to periodic functions
approximated by a Fourier series.  The steps are: approximate the
integrand with a Fourier series, with the coefficients found using the
fast Fourier transform; trivially integrate the series; evaluate the
integral at arbitrary points using Clenshaw
summation \citep{clenshaw55}.  \citet{trefethen14} review the
mathematical background for why this method gives such high accuracy.

In some cases, the integrands are almost singular, e.g., the term
$\sqrt{\kappa\cos^2\theta + \abs\mu}$ in the denominator of
Eq.~(\ref{ftheta}) leads to a sharp peak in the integrand when $\mu$
is very small, which in turn requires the inclusion of many terms in the
Fourier series.  This problem can be avoided by a change of variables
using
\begin{subequations} \label{F-am}
\begin{align}
x &= F(\phi, q), &
\phi &= \am(x, q),
\displaybreak[0]\\
\frac{\d x}{\d\phi} &= \frac1{\sqrt{1 - q^2\sin^2\phi}}, &
\frac{\d\phi}{\d x} &= \dn(x, q),
\end{align}
\end{subequations}
where $F(\phi, q)$ is the elliptic integral of the first kind, $\am(x,
q)$ is the Jacobi amplitude function, and $\dn(x, q)$ and $\cn(x, q)$
(used below) are Jacobi elliptic functions.  We adopt the notation
of \citep[Chaps.~19 \& 22]{dlmf10} except that, to avoid confusion with
$k$ defined in Eq.~(\ref{param-def}), we use $q$ to denote the modulus.

Substituting
\begin{equation}
\theta = \am\bigl(v, \sqrt{\kappa/(\kappa + \abs\mu)}\bigr) \label{vtx}
\end{equation}
in Eq.~(\ref{ftheta}) gives
\begin{equation}
f_\theta(\theta) = \int_0^{\am\theta}
\frac{\sqrt{1-\epsilon\kappa\cn^2v}}
{\sqrt{\kappa'+\kappa\cn^2v}\sqrt{\kappa+\abs\mu}}\,\d v,
\label{ftheta-p}
\end{equation}
where, for brevity's sake, we have omitted the modulus $q = \kappa/({\kappa
+ \abs\mu})$.  The same change of variables is made with
Eq.~(\ref{gtheta}).

In Eqs.~(\ref{fgstar}), we substitute
\begin{equation}
\psi = \am\bigl(u, \sqrt{(\kappa - \abs\mu)/\kappa}\bigr), \label{utx}
\end{equation}
for $\abs\mu$ small, to cancel the factor $\sqrt{\kappa\cos^2\psi
+ \abs\mu\sin^2\psi}$ in Eqs.~(\ref{fgstar}).  Finally, in
Eqs.~(\ref{gumb}) and (\ref{delta-f0}), we substitute $\theta
= \am(w, \sqrt\kappa)$ for $\kappa'$ small, to cancel the factor
$\sqrt{\kappa' + \kappa\cos^2\theta}$ in Eq.~(\ref{delta-f0}).

Even though the cost of computing $F(\phi, q)$ and $\am(x, q)$ is
small, we need only incur the cost when $q$ is sufficiently close to 1,
e.g., $q^2 > 7/8$.

There is another instance where the integrands are nearly singular,
namely for an almost flat ellipsoid with $c/b$ small, i.e., $ek$ close
to unity.  In this case, the factor $\sqrt{1-e^2k^2\cos^2\beta}$
appearing in Eqs.~(\ref{jacobi-comb}) has a sharp dip at $\beta = 0$.
The dip could be smoothed out by a suitable change of the variable of
integration, but that will take us too far afield.  The case where $c$
vanishes is discussed in Appendix \ref{app-billiards}

\section{The direct problem}

The direct problem, determining the position at a given distance along a
geodesic, is found by solving the nonlinear simultaneous equations
Eqs.~(\ref{comb-mod}).  This is accomplished using Newton's method in
two dimensions; details are given in Appendix \ref{newt2-sec}.  We need,
first, to specify the two independent variables to use.

For general geodesics $\gamma \ne 0$, the domains of $\psi$ and $\theta$
are unbounded.  It is preferable to use the new integration variables,
$u$ and $v$ defined by Eqs.~(\ref{utx}) and (\ref{vtx}), as the
independent variables instead of $\psi$ and $\theta$ (assuming a change
of variable was needed), because this results in smoother functions.

For umbilical geodesics, it is important to ``transform away'' the
singular behavior of $f_\psi$ and $f_\theta$ at the umbilics by using
\begin{equation}
u = \sinh^{-1}(k' \tan\psi),\qquad
v = \sinh^{-1}(k \tan\theta),\label{uumb}
\end{equation}
as the independent variables, so that the leading spherical
contributions Eqs.~(\ref{fgk00}) become
\begin{subequations}\label{fgk00u}
\begin{align}
\label{fk00u}
f_\psi(\psi; \epsilon=0, \mu=0) &=\frac{u}{kk'},\displaybreak[0]\\
\label{gk00u}
g_\psi(\psi; \epsilon=0, \mu=0) &= \tan^{-1}\frac{k\tanh u}{k'}.
\end{align}
\end{subequations}
In solving problems with umbilical geodesics, we keep track of transits
through umbilics so that we can make the restrictions,
$\abs\psi\le\half\pi$ and $\abs\theta\le\half\pi$.  The problem now maps
to the infinite domain in $(u,v)$ coordinates, as is the case for
general geodesics.

There are a few special cases in dealing with umbilical geodesics: the
geodesic lies on the median ellipse for which
$\delta \rightarrow \infty$ or the target point is an umbilic, in which
case $u$ and $v$ diverge.  The strategy in these cases is to let
$\delta$, $u$, or $v$ take on a large but finite value, so that, for
example, inverting Eq.~(\ref{uumb}) gives $\psi = \pm\half\pi$ with high
accuracy, but not so large that $\sinh u$ overflows.  I find that values
close to $-3 \log \epsilon$ suffice; here $\epsilon$ is the machine
epsilon, typically $2^{-52}$.  In this case, $\cos\psi =
k'/\sqrt{\sinh^2 u + k'^2}$ is nonzero (and similarly for $\cos\theta$),
allowing $\alpha$ to be determined from Eq.~(\ref{alpha-umb}).

Turning to the special case of biaxial ellipsoids, we treat them, as
much as possible, the same as triaxial ellipsoids.  This means a
slightly different treatment for meridional geodesics compared with
umbilical geodesics.  One other change was found to help in maintaining
accuracy.  Normally, $f_\psi(\psi; {\kappa = 1})$ is computed by
substituting $\kappa = 1$ into Eq.~(\ref{fstar}).  For triaxial
ellipsoids with small $\gamma$, we make the change of variables
Eq.~(\ref{utx}) to smooth out the integrand.  In the corresponding
situation for biaxial ellipsoids, it's better to determine
$f_\psi(\psi; {\kappa = 1})$ using Eqs.~(\ref{fdf-obl}) and (\ref{f10m}),
where the near-singular term is {\it subtracted} from the integrand and
integrated analytically.

\section{The inverse problem}\label{inverse}

\begin{figure}[tb]
\begin{center}
\includegraphics[scale=0.75]{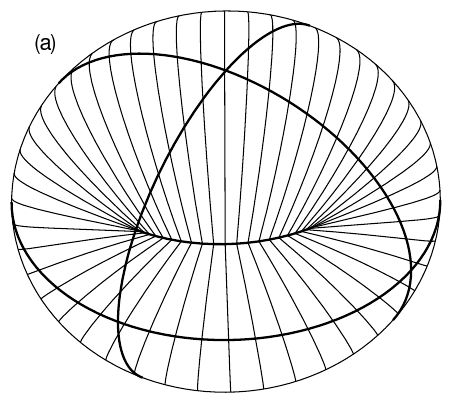}\\[2ex]
\includegraphics[scale=0.75]{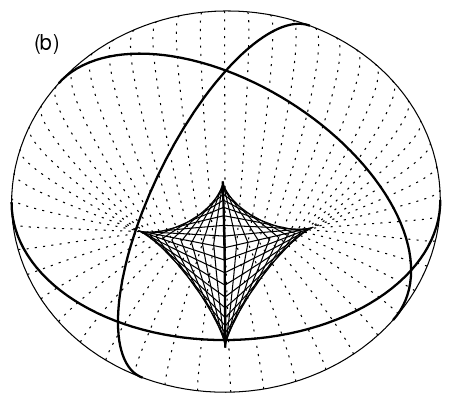}
\end{center}
\caption{\label{cut-figs}
Geodesics emanating from a single point on an ellipsoid.  The ellipsoid
and viewpoint are the same as Fig.~\ref{graticule}.  The geodesics start
at $\beta_1 = -34.46^\circ$, $\omega_1 = -149.94^\circ$, and the
azimuths $\alpha_1$ are multiples of $7.5^\circ$.  The geodetic
coordinates of the starting point are $\phi_1 = -40^\circ$, $\lambda_1 =
30^\circ-180^\circ$, the opposite of the viewing direction. Part (a)
shows the geodesics followed up to the points where they are no longer
the shortest geodesics.  The union of such points, the cut locus,
is shown as a heavy line and is a segment of the line of curvature
$\beta = -\beta_1$.  Part (b) shows the geodesics from (a) as dotted
lines, and these are continued (as solid lines) until they meet at
the line of curvature $\omega = \omega_1 + 180^\circ$ (shown as a heavy
line). }
\end{figure}%
\begin{figure}[tb]
\begin{center}
\includegraphics[scale=0.75]{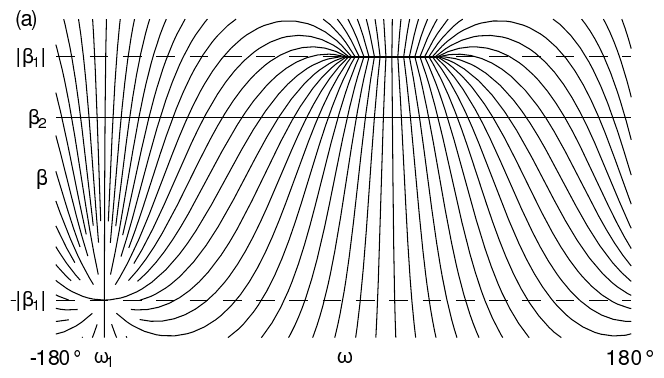}\\[2ex]
\includegraphics[scale=0.75]{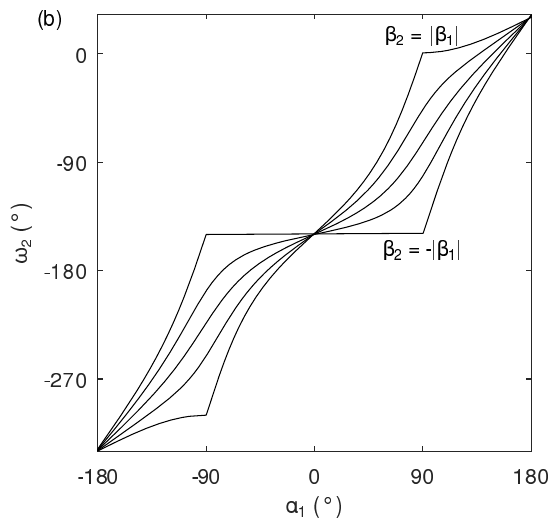}
\end{center}
\caption{\label{cut-unroll}
(a) The geodesics in Fig.~\ref{cut-figs}(a) displayed in a plate
carr\'ee (longitude-latitude) projection.  The geodesics all start at
$(\beta_1, \omega_1)$, with azimuths $\alpha_1$ which are multiples of
$7.5^\circ$, and are continued until they encounter the cut locus (shown
as a heavy line).  For a given $\beta_2 \in
[-\abs{\beta_1}, \abs{\beta_1}]$, the longitude $\omega_2$ is an
increasing function of $\alpha_1$.  (b) The mapping from $\alpha_1$ to
$\omega_2$ for the geodesics in (a).  The starting coordinates
$(\beta_1, \omega_1)$ are fixed, and curves of $\omega_2
= \omega_2^*(\alpha_1; {\beta_1, \omega_1, \beta_2})$ are shown for
$\beta_2 = [-1,-0.6,0,0.6,1]\times\abs{\beta_1}$.  The corresponding
plot for a biaxial ellipsoid is given by \citet[Fig.~4]{karney13}.
}
\end{figure}%
Crucial to solving the inverse problem is an understanding of the
properties of all the geodesics emanating from a single point
$(\beta_1, \omega_1)$.
\begin{itemize}
\item
If the starting point is an umbilic, all the lines meet at the opposite
umbilic at a distance $s_0$ given by Eq.~(\ref{umb-dist}).
\item
Otherwise, the first envelope of the geodesics is a four-pointed
astroid; see Fig.~\ref{cut-figs}(b).  Two of the cusps of the astroid
lie on $\beta = -\beta_1$, and the other two lie on $\omega = \omega_1
+ \pi$.  This is the so-called ``last geometric statement''
of \citet[\S6]{jacobi43}.
\item
All the geodesics intersect (or, in the case of $\alpha_1 = 0$ or $\pi$,
touch) the line $\omega = \omega_1 + \pi$.
\item
All the geodesics intersect (or, in the case of $\alpha_1 = \pm\half\pi$,
touch) the line $\beta = -\beta_1$.
\item
The two geodesics with azimuths $\pm\alpha_1$ first intersect on $\omega
= \omega_1 + \pi$, and their lengths to the point of intersection are
equal.
\item
The two geodesics with azimuths $\alpha_1$ and $\pi-\alpha_1$ first
intersect on $\beta = -\beta_1$, and their lengths to the point of
intersection are equal.
\end{itemize}
The last property defines the {\it cut locus} for $(\beta_1, \omega_1)$;
this is the locus of points where the geodesics cease to be shortest
geodesics.  This is a segment of a line of curvature $\beta = -\beta_1$;
see Fig.~\ref{cut-figs}(a).  This figure shows the shortest geodesic
between $(\beta_1, \omega_1)$ and {\it any} other point
$(\beta_2, \omega_2)$ on the ellipsoid.  Without loss of generality, we
take $\beta_1 \le 0$.  Then, for $\beta_2 \in
[-\abs{\beta_1}, \abs{\beta_1}]$, each geodesic intersects the line
$\beta = \beta_2$ exactly once.  For a given $\beta_2$, this defines a
monotonic mapping $\omega_2^*(\alpha_1; {\beta_1, \omega_1, \beta_2})$
of the circle of azimuths $\alpha_1$ to the circle of longitudes
$\omega_2$; see Fig.~\ref{cut-unroll}.  The mapping is continuous except
if $\beta_1 = 0-$ and $\beta_2 = 0$, where, for example, $\omega_2^*$
jumps from $\omega_1$, for $\alpha_1=\half\pi-$, to the conjugate
longitude, for $\alpha_1=\half\pi+$.

These properties show that the inverse problem can be solved using
techniques almost the same as those employed for a biaxial ellipsoid.
The first task is to treat those cases where both endpoints are on one
of the principal ellipses to determine whether the shortest geodesic
lies on that ellipse.

Starting with the median principal ellipse, we have the following
possibilities:
\begin{itemize}
\item
If the points are opposite umbilics, an arbitrary $\alpha_1$ may
be chosen.  Two of the shortest paths follow the median ellipse.
However, it is more useful to pick the geodesic which contains the point
$\beta = 0$, $\omega = \half\pi$; this gives finite and nonzero
values for $\tan\alpha_1$ and $\tan\alpha_2$, allowing other shortest
geodesics to be generated.
\item
Otherwise, if the shortest path on the ellipse crosses no more than a
single umbilic, the shortest geodesic lies on the median ellipse.  This
includes all cases where at least one of the endpoints is an umbilic.
\item
Otherwise, if the endpoints satisfy $\sin\omega_{1,2} = 0$ (the points
are near $X = \pm a$), the shortest geodesic is on the median ellipse.
\item
Otherwise, if the endpoints satisfy $\cos\beta_{1,2} = 0$ (the points
are near $Z = \pm c$), the shortest geodesic is on the median ellipse
only if there is no intervening conjugate point.  Otherwise, there are
two shortest paths, and one of them can be found using the general
method (given below) with the azimuth restricted to, e.g., $\alpha_1 \in
(-\half\pi, \half\pi)$, assuming that $\beta_1 = -\half\pi)$.
\end{itemize}

We turn now to the other principal ellipses.  The case where both points
are on the major ellipse, $\beta_{1,2} = 0$, is treated in the same way
as the last subcase for the median ellipse.  The shortest path is along
the ellipse, provided there is no intervening conjugate point.
Otherwise, there are two shortest paths, one of which has e.g.,
$\alpha_1 \in (-\half\pi, \half\pi)$, and this is found using the
general method.  If both points are on the minor principal ellipse, the
shortest path always follows the ellipse.

There is another instance where the azimuths can be found directly,
namely, if only one of the endpoints is an umbilic.  We then have
$\gamma = 0$, and the azimuth at the other point is given by
Eq.~(\ref{alpha-umb}), picking the signs of the sine and cosine
appropriately.

We now come to the general case where the shortest geodesic does not lie
on a principal ellipse and neither point is an umbilic.  The process
closely follows \citet[\S4]{karney13}.  Using symmetries, arrange that
$\beta_1 \le 0$ and $-\abs{\beta_1} \le \beta_2 \le \abs{\beta_1}$, the
situation depicted in Fig.~\ref{cut-unroll}.  Find the azimuth
$\alpha_1$ at point 1 which satisfies
\begin{equation}
\label{hybrid}
\omega_2^*(\alpha_1; \beta_1, \omega_1, \beta_2) = \omega_2 \pmod{2\pi};
\end{equation}
this is a one-dimensional root-finding problem.  Finally, determine the
length of the geodesic segment.

A few remarks are in order:
\begin{itemize}
\item
Finding the root $\alpha_1$ of Eq.~(\ref{hybrid}), which may require
several iterations, only requires consideration of
Eq.~(\ref{jacobi-mod}), which determines the course of the geodesic.
The calculation of the distance using Eq.~(\ref{ds-mod}) can be
postponed until $\alpha_1$ has been found.
\item
With the biaxial problem, \citet[Eq.~(6.5.1)]{helmert80} provided a
formula for the {\it reduced length} $m_{12}$, allowing conjugate points
to be found by the condition, $m_{12} = 0$.  I know of no corresponding
formula for $m_{12}$ for the triaxial case.  Nevertheless,
for geodesics with $\alpha_1 = \pm\half\pi$, the first conjugate point
is given by the condition that $\beta$
has completed half an oscillation, which in turn implies that $\psi$ has
advanced by $\pi$.  Equation~(\ref{jacobi-mod}) can be used to give the
value of $\omega$ at that point.  In the special case of a biaxial
ellipsoid, the conjugate point for a meridional geodesic is given by
Eq.~(\ref{meridconj}); this is only needed for solving inverse problems
on a prolate ellipsoid.
\item
In the biaxial case, we were able to solve Eq.~(\ref{hybrid}) using
Newton's method because the necessary derivative was given in terms of
$m_{12}$.  In the triaxial case, we do not have an expression for
$m_{12}$, so we resort to a simpler root-finding method.
\item
For biaxial ellipsoids, rotate the points about the axis of symmetry so
that $\omega_1 = 0$ (resp.~$\beta_1 = -\half\pi$) for oblate
(resp.~prolate) ellipsoids.  In this case, meridional geodesics are
handled by the logic for the median principal ellipse.
\item
The treatment of prolate ellipsoids differs from previous
work \citep{karney13}.  There, the generalization of the oblate case
entailed finding the intersection of a geodesic with a circle of
geodetic latitude, which corresponds to a circle of ellipsoidal
longitude.  In this triaxial treatment, we find the intersection with a
line of constant ellipsoidal latitude, which corresponds to a meridian
on a prolate ellipsoid.
\end{itemize}

The shortest path is unique unless:
\begin{itemize}
\item
The length of the geodesic vanishes $s_{12}=0$, in which case any
constant can be added to the azimuths.
\item
The points are opposite umbilics (this only applies for triaxial
ellipsoids, i.e., $k\ne0$ and $k'\ne0$).  In this case, $\alpha_1$ can
take on any value, and $\alpha_2$ needs to be adjusted to maintain the
value of $\tan\alpha_1 / \tan\alpha_2$.
\item
$\beta_1 + \beta_2 = 0$ and $\cos\alpha_1$ and $\cos\alpha_2$ have
opposite signs.  In this case, there is another shortest geodesic with
azimuths $\pi - \alpha_1$ and $\pi - \alpha_2$.
\end{itemize}
Any azimuth can be used for the shortest path connecting two opposite
poles on a biaxial ellipsoid or any two opposite points on a sphere.

There is an additional interesting property of geodesics: the geodesic
distance between the points $(\beta_1, \omega_1)$ and
$(\beta_2, \omega_2)$ equals that between $(\beta_2, \omega_1)$ and
$(\beta_1, \omega_2)$.  This is a consequence of evaluating the
integrals appearing in Eqs.~(\ref{jacobi-comb}) between the same limits
in the two cases.  We can state this another way: Consider a curvilinear
rectangle whose sides are lines of curvature; the diagonals of this
rectangle are equal.  This is known as Ivory's Lemma.

\section{Implementation and results}

An implementation of the solutions of the direct and inverse geodesic
problems is provided with version 2.7 of the C++ library,
GeographicLib \citep{geographiclib27}.  It's practically impossible to
exhaustively document the methods here; the reader is referred to the
code for details.  Here, I give an overview of the important aspects of
the code.

The class for performing the Fourier approximations to the integrands in
Jacobi's solution was inspired by the support for periodic functions
that was added to Chebfun \citep{wright15}.  The number of sample points
is successively doubled until convergence as defined by the ``chopping''
criterion given by \citet{aurentz17}.  Constructing a Fourier series for
the integral is simple.

I provide an optimized computation of the inverse of the integral.  This
uses Newton's method to compute a single value of the inverse.  During
the course of refining the Fourier series, a significant speedup is
achieved by using the current Fourier series for the inverse to provide
accurate starting guesses of the inverse at the new sample points.
The last (and most costly) rounds of Fourier refinement require only a
single Newton iteration for each sample point.  My initial expectation
was that this would be useful in cases where many waypoints needed to be
found.  In the event, I used the two-dimensional Newton's method as
described in Appendix \ref{newt2-sec} to solve the direct problem; this
obviates the need for finding the Fourier series for the inverse.

Turning to the solution of the inverse problem, I will focus on the
general case.  As described in Sec.~\ref{inverse}, this involves solving
Eq.~(\ref{hybrid}) for $\alpha_1$.  The solution of Eq.~(\ref{hybrid})
starts by finding $\omega_2^*$ for the four umbilical directions
(these all use the same $f_\psi$ and $f_\theta$), and these serve to
bracket the solution $\alpha_ 1$.  The root is found by the method given
by \citet{chandrupatla97}.

The solutions to the direct and inverse problems also return a
``geodesic line'' object.  This holds the four functions $f_\psi$,
$f_\theta$, $g_\psi$, and $g_\theta$, and the constants $\delta$ and
$s_1$; this allows waypoints along the geodesic to be computed
efficiently.

Testing for geodesics on a triaxial ellipsoid is about two orders of
magnitude more challenging than the biaxial case.  The shape of the
ellipsoid is specified by 2 parameters ($e$ and $k$) instead of just one
(the flattening), and the solution of the inverse problem depends on the
longitudes of both endpoints instead of just their difference.  So I
limited my initial testing to a single ellipsoid, Cayley's ellipsoid
scaled to the median semiaxis, i.e., $[a,b,c] = [ \sqrt2, 1, 1/\sqrt2
]$, with an emphasis on exploring all the different inverse problems
outlined in Sec.~\ref{inverse}.  The test set \citep{geod3test} contains
$500\,000$ geodesics, computed at high precision, with the coordinates
of the endpoints given as integer degrees.  My testing also included
other ellipsoids with $a/c = 2$, including oblate and prolate
ellipsoids.  In addition, I tested with spheres ($a = c$) with different
values of $k$, and with a triaxial model of the earth \citep{bursa93}.
In the following, I will only report results for the published dataset
for Cayley's ellipsoid.  These should be regarded as rough indications
of those likely to be obtained with ellipsoids which are not too
eccentric.

Averaging over the data in the test set, the mean number of Fourier
coefficients required to represent any of the functions $f_\psi$,
$f_\theta$, $g_\psi$, or $g_\theta$ is about 30.  The solution of the
two-dimensional equations for the direct geodesic problem using Newton's
method requires, on average, 5 iterations.  The solution of the inverse
method requires an average of 8 iterations of Chandrupatla's method.

These figures are for double precision with the error tolerance set to
machine precision, so that nearly full double-precision accuracy is
achieved.  Repeating the tests at higher precisions (64, 113, and 256
bits of precision, instead of 53 for double precision) shows that the
number of Fourier coefficients scales proportionally to the number of
bits.  Two-dimensional Newton's method for the direct problem enjoys the
expected quadratic convergence: the error is squared on each iteration.
The convergence is somewhat slower for the iterative solution for the
inverse problem; on each iteration, the logarithm of the error is
multiplied by about 7/4 (versus 2 for quadratic convergence).

When assessing the errors, it makes little sense to directly compare the
ellipsoidal coordinates and azimuths because these vary very rapidly
near the umbilics.  A better approach is to compare these quantities
expressed as cartesian positions and directions.  The differences are
given in ``units in the last place'' ($\mathrm{ulp}$), which I define to
be $b/2^{53}$ for the error in the position (this is also used for the
errors in the distance returned by the solution of the inverse problem)
and $1/2^{53}\,\mathrm{rad}$ for the errors in the direction.  For the
earth, $1\,\mathrm{ulp}$ corresponds to about $0.7\,\mathrm{nm}$.  The
errors, so quoted, will be approximately the same as for long double
precision (with $2^{53}$ replaced by $2^{64}$ to match the increase in
precision).

For the test set for Cayley's ellipsoid, the mean error in the position
and direction returned for the solution to the direct problem is
$5\,\mathrm{ulp}$ and $6\,\mathrm{ulp}$, respectively.  For the solution
to the inverse problem, the mean error in the distance is
$3\,\mathrm{ulp}$.  For this case, we do not compare the azimuths to the
test data, because, for example, there may be multiple allowed azimuths.
Instead, we demand consistency in the forward and backward direct
problems given by the inverse solution, measuring the discrepancies in
the positions and directions at the opposite endpoint; these are
$6\,\mathrm{ulp}$ for the positions and $7\,\mathrm{ulp}$ for the
directions.

These mean errors are impressively small.  However, for practical
applications, we need to quantify the maximum errors.  Using the test
data can only give a lower bound because the data offers rather sparse
coverage; in particular, it omits potentially problematic geodesics that
pass very close to umbilics.  Based just on the test set, the maximum
errors in the position and direction for the direct problem are
$160\,\mathrm{ulp}$ and $1500\,\mathrm{ulp}$.  The maximum error in the
distance for the inverse problem is $90\,\mathrm{ulp}$; the maximum
discrepancy in the position with the resulting direct problems is
$9000\,\mathrm{ulp}$.

These maximum errors are still reasonably small.  But more exhaustive
testing on this and other ellipsoids will surely uncover instances where
the errors are larger.  A reasonable course would be to assess the
errors in the context of a specific application.  This will have the
advantage of narrowing the parameters for the ellipsoids and will set a
definite limit on the acceptable errors.  If the errors are too large,
the present implementation can be used at higher precision to help track
down where the errors are creeping in.

The routines were timed on an Intel Core i7-9400 processor
($3$--$4.7\,\mathrm{GHz}$) with the code compiled with g++ and level 3
optimization.  Using the test data for Cayley's ellipsoid, the solution
of the direct problem takes on average $53\,\mu\mathrm s$; subsequent
waypoints on the geodesic line can be computed at a cost of
$7\,\mu\mathrm s$ per point.  The average time to solve the inverse
problem is $220\,\mu\mathrm s$.  This can be compared to the
corresponding times for finding geodesics on arbitrary biaxial
ellipsoids in terms of elliptic integrals \citep{karney-geod2}.  Taking
the flattening to be $f = \half$, the average times are $6\,\mu\mathrm
s$ for the direct solution, $3.5\,\mu\mathrm s$ for waypoints, and
$10\,\mu\mathrm s$ for the inverse solution.

An implementation of the coordinate conversions given in
Appendix \ref{app-conversion} is included in GeographicLib, as is sample
code for solving the ordinary differential equations for the direct
geodesic problem (Appendix \ref{app-ode}).

\section{Conclusions}

In this paper, I have described the implementation of Jacobi's solution
to the direct geodesic problem for a triaxial ellipsoid.  The method
involves using Fourier series to represent the integrands appearing in
Jacobi's solution, which allows the integrals to be evaluated very
accurately.  In addition, I show how the bisection method can be applied
to Newton's method in two dimensions, which allows the coupled system of
nonlinear equations for the direct problem to be solved efficiently.
The solution for the inverse problem follows the same basic method as
for biaxial ellipsoids.

The code is only about ten times slower than the much simpler case of
the biaxial ellipsoid.  I had also hoped to be able to say that the code
is only ten times less accurate than the biaxial case.  While the
average errors {\it do} meet this condition, there are cases where the
errors are substantially larger; this requires more study.
Nevertheless, a key goal has been met: if necessary, the algorithms can
be run with high-precision arithmetic, at a reasonable cost, to obtain
more accurate results.

An alternative method for solving geodesics is to integrate the ordinary
differential equations (ODEs) directly.  This is most easily carried out
in cartesian coordinates as advocated by \citet{panou19}.  Some data
on using this approach are given in Appendix \ref{app-ode}.  The
distinctions are as follows:
\begin{itemize}
\item
The code for solving the ODEs is considerably simpler, provided that a
good ``off-the-shelf'' library for ODEs is available.
\item
The ODEs only provide a solution for the direct geodesic problem.  It
{\it is} possible to extend this method to solve the inverse problem, but
this increases the complexity considerably.
\item
Jacobi's solution is somewhat more accurate for typical distances.  For
long geodesics, the accuracy of Jacobi's solution is maintained, while
the ODE solution progressively degrades.
\item
The ODE solution solves the direct problem somewhat faster than the
Jacobi solution for typical distances.  However, the CPU time for
Jacobi's solution is independent of distance, while it is proportional
to distance for the ODE solution.
\item
The ODE solution can be easily extended to compute the reduced length
and the geodesic scale; see Appendix \ref{closed-stab}.
\end{itemize}

\begin{appendix}
\numberwithin{equation}{section}

\section{Coordinates for points on the ellipsoid}\label{app-conversion}

We consider the coordinates for points on the surface of the ellipsoid.
(Extending the coordinate system to treat arbitrary points is considered
in Appendix \ref{app-conv3}.)  Besides ellipsoidal coordinates, three
other sets of coordinates used for triaxial ellipsoids are: geodetic
coordinates $(\phi, \lambda)$ defined by
\begin{subequations}
\begin{equation}
\hat{\mathbf U} =
[\cos\phi \cos\lambda, \cos\phi \sin\lambda, \sin\phi]^T;
\end{equation}
parametric coordinates $(\phi', \lambda')$ defined by
\begin{equation}
\mathbf R =
[a \cos\phi' \cos\lambda', b \cos\phi' \sin\lambda',
c \sin\phi']^T;
\end{equation}
and geocentric coordinates $(\phi'', \lambda'')$ defined by
\begin{equation}
\hat{\mathbf R} =
[\cos\phi'' \cos\lambda'', \cos\phi'' \sin\lambda'', \sin\phi'']^T.
\end{equation}
\end{subequations}
Explicit conversions between any of these coordinates and cartesian
coordinates form a common pattern.  We obtain cartesian coordinates from
geodetic coordinates with
\begin{subequations}
\begin{align}
\mathbf p &= [a^n \cos\phi \cos\lambda, b^n \cos\phi \sin\lambda,
c^n \sin\phi]^T,\\
\mathbf R &= \frac{\mathbf p}
{\sqrt{p_x^2/a^2 + p_y^2/b^2 + p_z^2/c^2}}, \label{geodtocart}
\end{align}
\end{subequations}
with $n = 2$.
The opposite conversion is given by
\begin{subequations}\label{carttogeod-all}
\begin{align}
\mathbf q &= [X/a^n, Y/b^n, Z/c^n]^T,\label{carttogeod}\displaybreak[0]\\
\phi &= \atanx{q_z}{\sqrt{q_x^2 + q_y^2}},\displaybreak[0]\\
\lambda &= \atanx{q_y}{q_x}.
\end{align}
\end{subequations}
The corresponding conversions for parametric and geocentric coordinates
are given by substituting $n = 1$ and $n = 0$, respectively, in place of
$n = 2$.  For the parametric conversion, Eq.~(\ref{geodtocart}) reduces
to $\mathbf R = \mathbf p$, while for the geocentric conversion,
Eq.~(\ref{carttogeod}) reduces to $\mathbf q = \mathbf R$.

Equation (\ref{cart}) defines the conversion from ellipsoidal to
cartesian coordinates.  This may be inverted with
\begin{subequations}\label{cartinv}
\begin{align}
\mathbf q &= [X/a, Y/b, Z/c]^T,\displaybreak[0]\\
s &= k^2 q_x^2 + (k^2 - k'^2) q_y^2 - k'^2  q_z^2, \displaybreak[0]\\
t &= \sqrt{ s^2 + 4 k^2 k'^2 q_y^2}, \displaybreak[0]\\
\cos\beta &= \begin{cases}
\displaystyle \frac{\sqrt{(t + s)/2}}{k}, &\text{if $s\ge 0$},\\
\abs{q_y/\sin\omega}, &\text{otherwise},
\end{cases} \displaybreak[0]\\
\sin\omega &=
\begin{cases}
\displaystyle \sign(q_y)\frac{\sqrt{(t - s)/2}}{k'}, &\text{if $s< 0$},\\
0, &\text{if $t = 0$},\\
q_y/\cos\beta, &\text{otherwise},
\end{cases} \displaybreak[0]\\
\sin\beta&=\frac{q_z}{\sqrt{k^2 + k'^2 \sin^2\omega}},
\displaybreak[0]\\
\cos\omega&=\frac{q_x}{\sqrt{k^2 \cos^2\beta + k'^2}}.
\end{align}
\end{subequations}

For each of the three coordinate systems---geodetic, parametric, and
geocentric---a meridian ellipse, defined as a line of constant
longitude, lies in a plane containing the $Z$ axis.  For geodetic
coordinates, this plane is defined by $Y/X = (a/b)^n \tan\lambda$, with
$n = 2$; substitute $n = 1$ or $n = 0$ for parametric or geocentric
coordinates.  In general, a ``circle of latitude'', a line of constant
latitude, only lies in a plane for the parametric latitude (and it is
not a circle for $a\ne b$).

Unlike ellipsoidal coordinates, none of these three coordinate systems
is orthogonal.  However, we can define an azimuth $\zeta$ with respect
to ``geodetic'' north, defined by $\partial \mathbf R / \partial \phi$;
this north direction is the same for all three systems, namely
\begin{equation}
\mathbf N' = [-X Z/c^2, -Y Z/c^2, X^2/a^2 + Y^2/b^2]^T,
\end{equation}
where the prime is used to distinguish $\mathbf N'$ from $\mathbf N$,
which is measured with respect to ellipsoidal coordinates.  We can now
convert between the cartesian direction $\mathbf V$ and either $\alpha$
or $\zeta$.

\section{Geodetic coordinates for arbitrary points}\label{app-conv3}

Geodetic and ellipsoidal coordinates have natural extensions to
arbitrary points in three dimensions.  Geodetic coordinates are
generalized by giving the height normal to the ellipsoid.  Thus, a
position is given by
\begin{equation}\label{geodetic}
   \mathbf R = \mathbf R_0 + h \hat{\mathbf U}(\mathbf R_0),
\end{equation}
where $\mathbf R_0$ is the closest point on the ellipsoid and $h$ is the
height.  The full geodetic coordinates are then given by
$(\phi, \lambda, h)$.

The extension of ellipsoidal coordinates to three dimensions
places an arbitrary point on a confocal ellipsoid defined by
\begin{equation}\label{udef}
   \frac{X^2}{u^2 + l_a^2} +
   \frac{Y^2}{u^2 + l_b^2} + \frac{Z^2}{u^2} - 1 = 0,
\end{equation}
where $l_a = \sqrt{a^2 - c^2}$ and $l_b = \sqrt{b^2 - c^2}$ are linear
eccentricities and $u$ is its minor semiaxis.  The full ellipsoidal
coordinates are $(\beta, \omega, u)$, and the conversion for these to
cartesian coordinates is given by Eq.~(\ref{cart}), replacing $(a, b,
c)$ by $(\sqrt{u^2 + l_a^2}, \sqrt{u^2 + l_b^2}, u)$.

We have handled the conversion from $(\phi, \lambda, h)$ and
$(\beta, \omega, u)$ to cartesian.  Let us address the reverse
operation, starting with the conversion of cartesian coordinates to
geodetic.  This is a standard problem covered, for example,
by \citet[\S76]{bell12}.  The solution is given by finding the largest
root $p$ of
\begin{equation}\label{geodh}
   f(p) = \biggl(\frac{a X}{p + l_a^2}\biggr)^2 +
   \biggl(\frac{b Y}{p + l_b^2}\biggr)^2 +
   \biggl(\frac{c Z}{p}\biggr)^2 - 1 = 0.
\end{equation}
Then we have
\begin{subequations}
\begin{align}
   \mathbf R_0 &= \biggl(
   \frac{a^2 X}{p + l_a^2},
   \frac{b^2 Y}{p + l_b^2},
   \frac{c^2 Z}{p} \biggr)^T, \label{R0-eq}\displaybreak[0]\\
   h &= \hat{\mathbf U}(\mathbf R_0) \cdot (\mathbf R - \mathbf R_0)
   = (p - c^2)\, U(\mathbf R_0),
\end{align}
\end{subequations}
and $(\phi, \lambda)$ are given by Eqs.~(\ref{carttogeod-all}).

\citet{ligas12} uses Newton's method to find the root of
Eq.~(\ref{geodh}); however, with his choice of starting guess, this
sometimes fails to converge.  \citet{panou22} cure this defect by using
the bisection method to find the root.  This is guaranteed to converge,
but at a high computational cost.  Alternatively, \citet{diaztoca20} use
Newton's method to find the largest root of a sixth-order polynomial
obtained by converting Eq.~(\ref{geodh}) to a rational expression;
however, \citet{panou22} report that this also fails to converge in some
instances.

It turns out we can easily fix the problems with Newton's method applied
to Eq.~(\ref{geodh}).  First of all, note that $f(p)$ has positive
double poles at $p = 0$, $-l_b^2$, and $-l_a^2$ and that
$f(p) \rightarrow -1$ for $p \rightarrow \pm\infty$.  (For now, we
assume that $(X, Y, Z)$ are all nonzero.).  Therefore, $f(p)=0$ has a
unique root for $p \in (0, \infty)$.  In this region, $\d f(p)/\d p < 0$
and $\d^2 f(p)/\d p^2 > 0$, and, as a consequence, picking a starting
guess for Newton's method between $p = 0$ and the actual root is
guaranteed to converge.

To obtain a reasonably tight bound on the root, we note that
\begin{subequations}
\begin{align}
   f(p) &\le \frac{c^2 Z^2}{p^2} - 1, \displaybreak[0]\\
   f(p) &\le \frac{b^2 Y^2 + c^2 Z^2}{(p + l_b^2)^2} - 1,
   \displaybreak[0]\\
   f(p) &\le \frac{a^2 X^2 + b^2 Y^2 + c^2 Z^2}{(p + l_a^2)^2} - 1,
   \displaybreak[0]\\
   f(p) &\ge \frac{a^2 X^2 + b^2 Y^2 + c^2 Z^2}{p^2} - 1.
\end{align}
\end{subequations}
Because $\d f(p)/\d p < 0$ for $p > 0$, this leads to bounds on
the positive root, $p_{\mathrm{min}} \le p \le p_{\mathrm{max}}$,
where
\begin{subequations}
\begin{align}
   p_{\mathrm{min}} &= \max\Bigl( c\abs Z,
   \sqrt{b^2 Y^2 + c^2 Z^2} - l_b^2,\notag\\
&\qquad\qquad \sqrt{a^2 X^2 + b^2 Y^2 + c^2 Z^2} - l_a^2\Bigr),
\displaybreak[0]\\
   p_{\mathrm{max}} &= \sqrt{a^2 X^2 + b^2 Y^2 + c^2 Z^2}.
\end{align}
\end{subequations}
In implementing Newton's method, we neglect any term in the definition
of $f(p)$ if its numerator vanishes (even though the denominator might
also vanish).

Provided that $f(p_{\mathrm{min}}) > 0$, we can start Newton's method
with $p_0 = p_{\mathrm{min}}$, and this is guaranteed to converge to the
root from below.  If $f(p_{\mathrm{min}}) \le 0$, which can only happen
if $Z = 0$, the required solution is $p = 0$.  In this case, the
expression for $\mathbf R_0$ is indeterminate (at least one of the
components in Eq.~(\ref{R0-eq}) involves division by $0$), and we
proceed as follows:
\begin{itemize}
\item If $X_0$ is indeterminate, set $X_0 = 0$ (this
  can only happen with $X = 0$ on a sphere).
\item If $Y_0$ is indeterminate, set $Y_0 = 0$ (this
  can only happen with $Y = 0$ on an oblate spheroid).
\item Finally, set $Z_0 = \pm c \sqrt{1 - X^2/a^2 - Y^2/b^2}$.
\end{itemize}
This prescription obviates the need to enumerate and treat various
subcases as \citet{diaztoca20,panou22} do.

Turning to the question of converting a cartesian position into
ellipsoidal coordinates, we need to find the largest value of $u$ that
solves Eq.~(\ref{udef}).  Writing $q = u^2$, this becomes the task of
finding the largest root of
\begin{equation}\label{qdef}
   g(q) = \frac{X^2}{q + l_a^2} +
   \frac{Y^2}{q + l_b^2} + \frac{Z^2}q - 1 = 0.
\end{equation}
The structure of $g(q)$ resembles that of $f(p)$, Eq.~(\ref{geodh}).
Since $g(q)$ has 3 simple poles with positive coefficients, there are
three real roots, and, because the rightmost pole is at $q = 0$ and
because $g(q\rightarrow\infty) = -1$, just one of them is positive.  As
before, bounds can be put on this root $q_{\mathrm{min}} \le q \le
q_{\mathrm{max}}$, where
\begin{subequations}
\begin{align}
   q_{\mathrm{min}} &= \max(Z^2,
   Y^2 + Z^2 - l_b^2,
   X^2 + Y^2 + Z^2 - l_a^2), \displaybreak[0]\\
   q_{\mathrm{max}} &= X^2 + Y^2 + Z^2.
\end{align}
\end{subequations}

Provided that $g(q_{\mathrm{min}}) > 0$, we start Newton's method with
$q_0 = q_{\mathrm{min}}$, and this converges to the root from below.  If
$g(q_{\mathrm{min}}) \le 0$ (which can only happen if $Z = 0$), the
required solution is $q = u = 0$.

Of course, we can expand $g(q)$ to obtain a cubic polynomial
in $q$, which can be solved analytically.  This method is used by
\citet{panou21}.  The solution may be subject to unacceptable
roundoff error, but it can be refined by using it as the starting point,
$q_0$, for Newton's method (which will converge in one or two
iterations).  In this case, if $g(q_0) < 0$, $q_1$ should be replaced by
$\max(q_1, q_{\mathrm{min}})$.

Having determined $u = \sqrt q$, $\beta$ and $\omega$ may be found by
applying Eqs.~(\ref{cartinv}) to the confocal
ellipsoid.

\section{Newton with bisection in two dimensions} \label{newt2-sec}

\begin{figure}[tb]
\begin{center}
\includegraphics[scale=0.75]{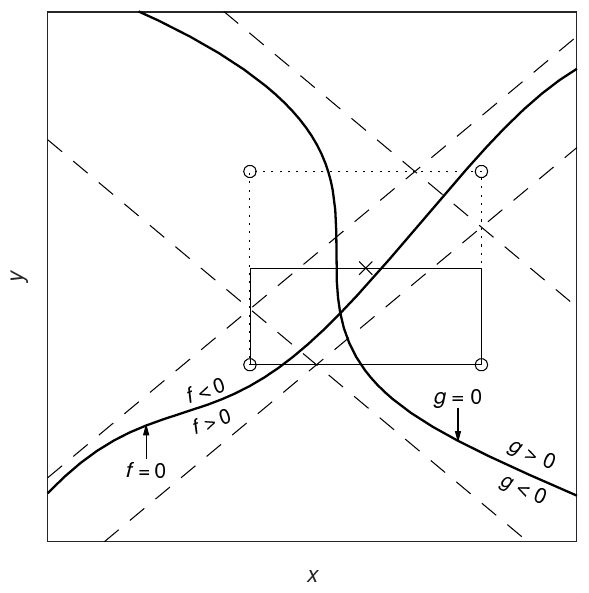}
\end{center}
\caption{\label{newt2-fig}
Bisection in two dimensions.  The heavy curves show $f(x,y) = 0$ and
$g(x,y) = 0$.  These curves are constrained to lie within the dashed
lines, and thus the solution to $f = g = 0$ must lie within the
parallelogram formed by these lines.  The rectangles show bounding boxes
for the solution, as explained in the text. }
\end{figure}%
Let us consider the solution of the coupled nonlinear equations
\begin{subequations}\label{2d-eqs}
\begin{align}
f(x, y) &= f_x(x) - f_y(y) - f_0 = 0, \label{2d-eqa}\displaybreak[0]\\
g(x, y) &= g_x(x) + g_y(y) - g_0 = 0. \label{2d-eqb}
\end{align}
\end{subequations}
These are Eqs.~(\ref{comb-mod}) with $x$ standing for $\psi$ or $u$ and
$y$ standing for $\theta$ or $v$.  Each of $f_x$ and $f_y$ is an
increasing function of its arguments, while $g_x$ and $g_y$ are
non-decreasing functions.  There is a unique solution to these equations
for given finite $f_0$ and $g_0$, with the proviso, for umbilical
geodesics, that $2b\abs{g_0}$ is less than or equal to the distance
between opposite umbilics.

Each of these functions consists of a secular term proportional to $x$
or $y$ and a bounded variation about this term.  The slope of the
secular part is positive for $f_x$ and $f_y$ and non-negative for $g_x$
and $g_y$, and it is straightforward to place an upper bound on the
variations (e.g., by summing the magnitudes of the coefficients of the
oscillating terms in the Fourier representations).  As a consequence, we
can bound the solution to a parallelogram in $(x,y)$ space as shown in
Fig.~\ref{newt2-fig}.

One approach to solving the two-dimensional system is to regard, say,
$y$ as the control variable and to write $x = f_x^{-1}(f_y(y) + f_0)$.
Here, $x$ is found by an inner invocation of Newton's method.  The $g$
equation is solved by an outer invocation.  This method is effective but
typically requires many function evaluations.

An alternative is to use Newton's method in two dimensions.  This is a
straightforward generalization of the normal one-dimensional method,
with the reciprocal of the derivative replaced by the inverse of the
Jacobian for the system.  The parallelogram bound on the solution can be
used to estimate a starting point and to detect when the method goes
awry.  However, in general, it is difficult to establish ever-narrowing
bounds on the solution so that a bisection step will provide a good
guess, allowing Newton's method to be continued.

It turns out that for the class of two-dimensional root-finding problems
given by Eq.~(\ref{2d-eqs}), it is quite easy to establish an
axis-aligned bounding
rectangle, $x \in [x_a, x_b]$ and $y \in [y_a, y_b]$.  This starts by
including all of the initial bounding parallelogram---a much looser
bound.  On each iteration, update one edge of the bounding rectangle
depending on the signs of $f$ and $g$.  Thus, in the example shown in
Fig.~\ref{newt2-fig}, the corners of the initial bounding box are marked
by circles, and the initial guess, marked by a cross, is at its center.
At this point, we find that $f<0$ and $g>0$.  From the figure, it is
clear that the desired solution lies below this point, and therefore, we
update $y_b$, giving a new bounding box shown as the light rectangle.
This is possible because of the constraints on the slopes of the curves
$f=0$ and $g=0$, which, in turn, result from the form of these functions
in Eqs.~(\ref{2d-eqs}).  With other sign combinations for $f$ and $g$,
the other edges of the bounding rectangle can be updated, and in the
case where either $f$ or $g$ vanishes, two edges can be updated.

There are several refinements possible.  First of all, the method
depends on the monotonicity of $f_x$, $f_y$, $g_x$, and $g_y$, and, in a
numerical context, this is not assured.  We remedy this by maintaining a
list of all the values of $x \in [x_a, x_b]$ encountered so far and the
associated values of $f_x(x)$ and $g_x(x)$ (and correspondingly for
$y$).  When a new value of $x$ is inserted, we ``clamp'' the value of
$f_x(x)$ and $g_x(x)$ to those of the neighboring values.

When a Newton iteration falls outside the bounding box or if the method
is converging too slowly, a new starting value is given by the values of
$x$ and $y$ that are the midpoints of the largest gaps in their
respective lists.

We can be more aggressive in updating the bounding rectangle.  Whenever
a new $x$ is added to the $x$ list, we update the bounds by checking the
signs of $f$ and $g$ with the new $x$ and all the previous $y$ values
(and also for each new $y$).  This prevents the lists for $x$ and $y$
from growing very long.

Sometimes the procedure given in the previous paragraph leads to a
violation of the constraints, which are obvious from
Fig.~(\ref{newt2-fig}): $f(x_a, y_b) \le 0$, $f(x_b, y_a) \ge 0$,
$g(x_a, y_a) \le 0$, and $g(x_b, y_b) \ge 0$.  This can happen, for
example, if $g = 0$ at several of the intersections of the $x$ and $y$
lists.  In this case, we update the bounds with {\it only} the new value
of $(x, y)$.

We also apply this method in the biaxial case for which $g_y(y) = 0$.
This is a degenerate case because it reduces to two one-dimensional
root-finding problems: use the $g$ equation Eq.~(\ref{2d-eqb}) to find
$x$ and then, with this value of $x$, use the $f$ equation
Eq.~(\ref{2d-eqa}) to find $y$.  The two-dimensional solution described
here sometimes suffers from poor convergence because the value of $x$
oscillates between two consecutive floating-point numbers, neither of
which exactly satisfies the $g$ equation.  In this case, we adjust $g_0$
to allow the $g$ equation to be satisfied exactly, and the
two-dimensional solution proceeds to solve the $f$ equation in the same
manner as the one-dimensional solution.

\section{The elliptical billiard problem}\label{app-billiards}

\begin{figure}[tb]
\begin{center}
\includegraphics[scale=0.75]{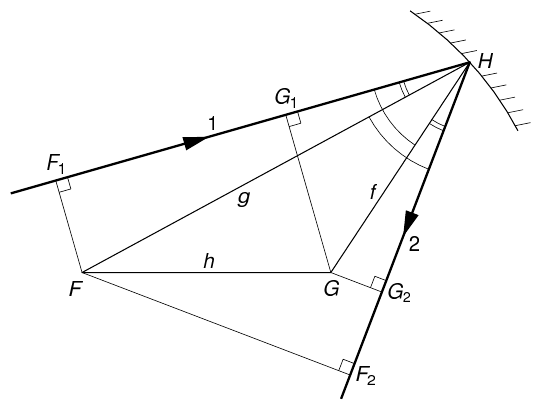}
\end{center}
\caption{\label{similar-fig}
A billiard with trajectory marked $1$ and $2$ (shown as heavy lines)
bouncing off the wall of the table at $H$.  The foci of the elliptical
wall are $F$ and $G$, and standard conventions are used to indicate
equal angles.  The right triangles $HFF_1$ and $HGG_2$ are similar,
which gives $\abs{HF}/\abs{HG} = \abs{FF_1}/\abs{GG_2}$.  Also $HFF_2$
and $HGG_1$ are similar, which gives $\abs{HF}/\abs{HG}
= \abs{FF_2}/\abs{GG_1}$.  The sides of the triangle $FGH$ are $f$, $g$,
and $h$. }
\end{figure}%
\begin{figure*}[tb]
\begin{center}
\includegraphics[scale=0.75]{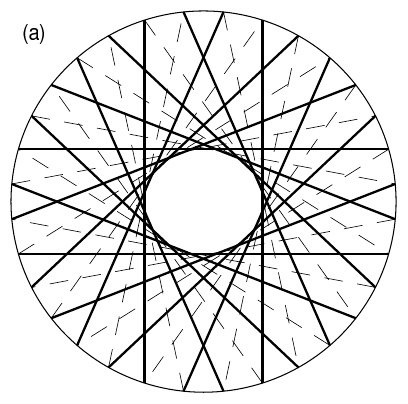}\hspace{1em}
\includegraphics[scale=0.75]{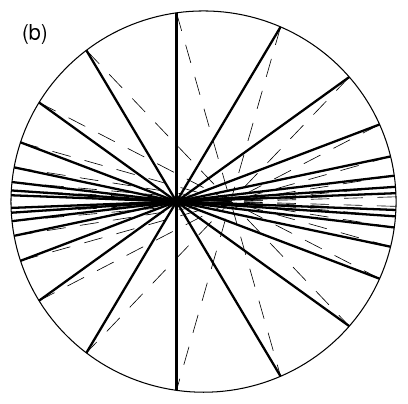}\hspace{1em}
\includegraphics[scale=0.75]{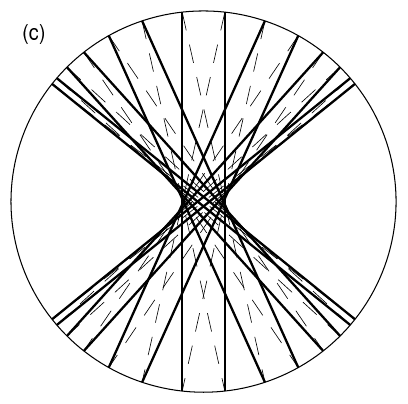}\hspace{1em}
\end{center}
\caption{\label{billiard-fig}
Samples of geodesics for an ellipsoid with $a = 1.01$, $b = 1$, and
$c = 0$, i.e., an elliptical disk.  The geodesics on the upper
(resp.\ lower) face of the ellipse are shown as solid (resp.\ dashed)
lines.  Parts (a), (b), and (c) show geodesics with $\gamma > 0$,
$\gamma = 0$, and $\gamma < 0$, respectively. }
\end{figure*}%
In the limit $c \rightarrow 0$, i.e., $ek\rightarrow 1$, the ellipsoid
is flattened to an elliptical disc, and the geodesic problem reduces to
the problem of a ball bouncing off the walls of an elliptical billiard
table.  To make the correspondence exact, we would further (fancifully)
stipulate that on each bounce the ball switches between the top and
bottom of the table.  In this case, the factor
$\sqrt{1-e^2k^2\cos^2\beta}$ appearing in Eqs.~(\ref{jacobi-comb})
reduces to $\abs{\sin\beta}$, and the integrand can be well represented
by a Fourier series, provided we split the integral up into pieces
depending on the sign of $\sin\beta$; this is the same as the procedure
used to obtain the equations for umbilical geodesics Eqs.~(\ref{fgumb}).

But this is a very roundabout way to obtain the trajectory of the
billiard ball, which can be obtained by elementary trigonometry.  In
this case, the simple closed geodesics are: rolling around the edge of
the table; following the minor axis of the ellipse (both of these are
stable); and bouncing between the foci of the ellipse (this is
unstable).  These cases correspond, respectively, to geodesics following
the major, minor, and median principal ellipses for the ellipsoid.

The conserved quantity $\gamma$ Eq.~(\ref{gamma-def}) has a simple
interpretation: it is proportional to the product of the angular momenta
about the two foci of the ellipse.  This conservation law is proved
using the similar triangles shown in Fig.~\ref{similar-fig}.  Because
the speed of the ball is constant, its angular momentum on trajectory
$1$ about the focus $F$ is proportional to the perpendicular distance
$\abs{FF_1}$, etc.  Combining the relationships for the similar
triangles in Fig.~\ref{similar-fig}, we obtain
\begin{equation}
\abs{FF_1}\abs{GG_1} = \abs{FF_2}\abs{GG_2},
\end{equation}
which establishes that the product of the angular momenta is conserved
on a bounce.  Typical paths are shown in Fig.~\ref{billiard-fig};
compare these with the geodesics shown in Fig.~\ref{sample-fig}.  For
$\gamma > 0$, Fig.~\ref{billiard-fig}(a) (resp.~$\gamma < 0$,
Fig.~\ref{billiard-fig}(c)), the path of the billiard ball is tangent to
a confocal ellipse (resp.~hyperbola).

The instability of the path connecting the two foci, $F$ and $G$, is
easily established by considering the triangle $FGH$ in
Fig.~\ref{similar-fig}.  We adopt the usual nomenclature for triangles,
where $F$, $G$, and $H$ measure the interior angles and $f$, $g$, and $h$
are the lengths of the opposite sides.  One of Mollweide's formulas for
a triangle gives
\begin{align}
\tan\tfrac12 F \tan \tfrac12 G &= \frac{(f+g) - h}{(f+g) + h}
=\frac{a-\sqrt{a^2-b^2}}{a+\sqrt{a^2-b^2}} \notag\\
&=\exp(-\Delta),\label{moll}
\end{align}
where we have substituted $f + g = 2a$ and $h = 2\sqrt{a^2-b^2}$, and
$\Delta$ is given by either of Eqs.~(\ref{Delta}) or (\ref{Delta-hart})
with $c = 0$.  On successive bounces, the tangents of the half-angles at
the left focus form a geometric progression, increasing by a factor
$\exp(2\Delta)$ on each passage through the left focus.  If the path is
followed in the forward direction, it will coincide with the major axis
of the ellipse, while if followed in the reverse direction, it will lie
on the major axis in the opposite sense.  This is illustrated in
Fig.~\ref{billiard-fig}(b).

\section{Ordinary differential equations for geodesics}
\label{app-ode}

\citet[\S8]{laplace99} showed that a particle constrained to move on
a surface but subject to no other forces follows a geodesic.
The centrifugal acceleration of the particle is
$-(V^2/R_c)\hat{\mathbf U}$, where $R_c$ is the radius of curvature in
the direction of the velocity $\mathbf V$.  We will take the speed $V$ to be
unity (and, of course, the speed is a constant in this problem); thus,
time can be replaced by $s$, the displacement along the geodesic, as the
independent variable.  The differential equations for the geodesic are
\begin{subequations} \label{odes}
\begin{align}
\d\mathbf R/\d s &= \mathbf V, \label{oder} \displaybreak[0]\\
\d\mathbf V/\d s &= \mathbf A, \label{odev}
\end{align}
where
\begin{equation}
\mathbf A = -\frac{\mathbf U}{U^2}
   \biggl( \frac{V_x^2}{a^2} + \frac{V_y^2}{b^2} + \frac{V_z^2}{c^2} \biggr).
\end{equation}
\end{subequations}
This expression for the acceleration $\mathbf A$ is obtained by
computing the deviation of the particle from the ellipsoid if $\mathbf
V$ is constant; the acceleration necessary to maintain the particle on
the ellipsoid immediately follows.  \citet{panou13} solves these
equations in ellipsoidal coordinates, but this leads to a badly behaved
system because of the singular behavior of these coordinates near
umbilics.  A better approach, adopted by \citet{panou19}, is to express
$\mathbf R$ and $\mathbf V$ in cartesian coordinates, because there are
no singularities in this representation.

\citet{panou19} integrated the system using a 4th-order Runge-Kutta
scheme.  Because this is a relatively low-order method, it's necessary
to use a small step size to control the truncation error.
Unfortunately, because a large number of steps are required, this might
give an unacceptably large roundoff error.  They mitigated this by using
``long double'' precision (with 64 bits in the fraction as opposed to 53
bits for standard double precision).

A better approach is to use a high-order integration method.  Such
methods typically adjust the step size automatically to obtain the
desired accuracy.  After some experimentation, I found satisfactory
tools to integrate Eqs.~(\ref{odes}), as follows:
\begin{itemize}
\item
Octave's \verb|ode45|: This is an implementation of the Dormand-Prince
method.
\item
MATLAB's \verb|ode89|: A high-order Runge-Kutta method.
\item
Boost's \verb|bulirsch_stoer|: This Bulirsch-Stoer
method uses Richardson extrapolation to obtain an accurate solution, and
it can be used with floating-point numbers of any precision.
\item
Boost's \verb|bulirsch_stoer_dense_out|: This is a
variant of the previous method providing ``dense output,'' i.e.,
accurately interpolated results within a step.
\end{itemize}
All these methods allow the direct geodesic problem to be solved about
as accurately and about as fast as using Jacobi's method described in
the body of this paper.  For example, applying the Bulirsch-Stoer method
to the test set for Cayley's ellipsoid, I find that the average error in
the position at point 2 is $120\,\mathrm{ulp}$, and the average CPU time
is $15\,\mu\mathrm s$.  On average, 10 integration steps are required.
However, the length of the geodesics in the test set is bounded---they
are all ``shortest geodesics.''  For longer geodesics, the CPU time will
scale linearly with distance and the accuracy will degrade.

These errors in the solution given by integrating the ordinary
differential equations are an impediment to solving the inverse problem.
This depends sensitively on certain properties of the solution, e.g.,
that a geodesic leaving an umbilic intersects the opposite umbilic.  I
have stitched up a solution for the inverse problem in the Octave/MATLAB
version of GeographicLib by tracking the solution in ellipsoidal
coordinates as the solution in cartesian coordinates
unfolds \citep{geographiclib-octave25}.  But the code is somewhat
messy, because it involves repeated coordinate conversions and has to
work around the small, but inevitable, errors in the solution of the
direct problem.

\section{The stability of closed geodesics}\label{closed-stab}

The stability of a geodesic is determined by
\begin{equation}
t'' = -K t, \label{redlen}
\end{equation}
where $t$ is the infinitesimal separation of a geodesic from a
reference geodesic, prime indicates differentiation with respect to
$s$ (i.e., $t'' = \d^2 t/\d s^2$), and
\begin{align}
K &= \frac{a^2 c^2 / b^6}
{(1+e^2k'^2\sin^2\omega)^2(1-e^2k^2\cos^2\beta)^2}\notag\\
&= \frac1{a^2b^2c^2 U^4}\label{curvature}
\end{align}
is the Gaussian curvature.  The first form of $K$ is given
by \citet[\S3.5.11]{klingenberg82}, and the second is obtained by
converting it to cartesian coordinates.  Equation (\ref{redlen}) is
solved with two sets of initial conditions,
\begin{itemize}
\item
$t(0) = 1$, $t'(0) = 0$, and then $t = M_{12}$ is the forward geodesic
scale,
\item
$t(0) = 0$, $t'(0) = 1$, and then $t = m_{12}$ is the reduced length,
and $t' = M_{21}$ is the reverse geodesic scale.
\end{itemize}
Some of the properties of $m_{12}$, $M_{12}$, and $M_{21}$ are given
in \citet[\S3]{karney13}.  I solve for these quantities by supplementing
the ordinary differential equations in Appendix \ref{app-ode} with
Eq.~(\ref{redlen}).  This gives a system of ten first-order differential
equations, six for the geodesic and four for $m_{12}$ and $M_{12}$.

For a closed geodesic on one of the principal ellipses, $K(s)$ is a
periodic function with period $s_0$, one half of the perimeter of the
ellipse.  Thus, Eq.~(\ref{redlen}) is an example of Hill's
equation \citep[\S28.29]{dlmf10}, which can be solved using Floquet's
theorem.  To determine the behavior of the equation for large $s$, it
suffices to solve it over one period and form the matrix,
\begin{equation}
\mathcal M =
\begin{bmatrix}
M_{12} & m_{12} \\
M'_{12} & m'_{12}
\end{bmatrix}
=
\begin{bmatrix}
M_{12} & m_{12} \\
{\displaystyle-\frac{1-M_{12} M_{21}}{m_{12}}} & M_{21}
\end{bmatrix},
\end{equation}
where all the terms in the matrix are evaluated at $s = s_0$.  The
Wronskian of this system is unity, $\det \mathcal M = 1$.  If the
initial conditions are $t(0) = W_0$ and $t'(0) = w_0$, then, after an
integer $l$ periods, we have $t(l s_0) = W_l$ and $t'(l s_0) = w_l$
where
\begin{equation}
\begin{bmatrix}W_l\\w_l\end{bmatrix}
= \mathcal M^l
\begin{bmatrix}W_0\\w_0\end{bmatrix}.
\end{equation}
The stability of the system is determined by the eigenvalues of
$\mathcal M$.  Writing $\tr \mathcal M = M_{12} + M_{21} = 2M$, the
eigenvalues are
\begin{equation}
\lambda_{1,2} = M \pm \sqrt{M^2 - 1},
\end{equation}
with $\lambda_1\lambda_2 = 1$.  We distinguish four cases:
\begin{itemize}
\item
if $\abs M > 1$, one of $\abs{\lambda_{1,2}}$ is greater than unity, and
the solution is exponentially unstable;
\item
if $\abs M < 1$, $\abs{\lambda_{1,2}}$ are both unity, and the solution
is stable (bounded oscillations);
\item
if $\abs M = 1$ and the off-diagonal terms of $\mathcal M$ vanish,
the solution is stable;
\item
if $\abs M = 1$ and at least one off-diagonal term of $\mathcal M$ is
nonzero, the solution is linearly unstable.
\end{itemize}
Computing $M$ for the principal ellipses of Cayley's ellipsoid, we have
$M = -0.3634$, $-1.6399$, and $0.4274$ for the minor, median, and major
ellipses.  This confirms that the median ellipse is exponentially
unstable, while the other two are stable.  Examples where $\abs M = 1$
are: great circles on a sphere (stable), the equator for an oblate
ellipsoid with integer $b/c$ (stable), and meridian ellipses on biaxial
ellipsoids (linearly unstable).

The rate of instability for the median ellipse is reflected in the
quantity $\Delta$ introduced in Sec.~\ref{umb-sec}.  However, we can
generalize that result to apply to any of the principal ellipses.  In
Hart's equation for $\Delta$ Eq.~(\ref{Delta-hart}), $a$ and $c$, the
semiaxes of the median ellipse, appear symmetrically, while $b$ occupies
a privileged position.  By making suitable exchanges between $a$, $b$,
and $c$, we obtain the corresponding values of $\Delta$ for all three
principal ellipses, giving $\Delta = 1.1989 \, i$, $1.0783$, and
$2.0125 \, i$ (with $i = \sqrt{-1}$).  We now find that $M$ and $\Delta$
are related by $M = -\cosh \Delta$.

\end{appendix}

\paragraph*{Data availability}
Test data is available at \citet{geod3test}.  Additional test data is
available upon request.
\paragraph*{Code availability}
The C++ implementation of the algorithms given in this paper is given in
GeographicLib, version 2.7, available at
\url{https://github.com/geographiclib/geographiclib/releases/tag/r2.7}.
This version is archived at \url{https://doi.org/10.5281/zenodo.18011538}.

\bibliography{geod}
\end{document}